\documentclass[10pt,twocolumn,letterpaper]{article}

\usepackage[pagenumbers]{cvpr} 

\usepackage{graphicx}
\usepackage{amsmath}
\usepackage{amssymb}
\usepackage{booktabs}
\usepackage{pifont}

\usepackage{bbding}
\usepackage{soul}
\usepackage{bibunits}
\usepackage{bm}
\usepackage{array}
\usepackage[table]{xcolor}
\usepackage{pdflscape}
\usepackage{makecell}
\usepackage{framed,multirow}
\usepackage{threeparttable}
\usepackage{amssymb}
\usepackage{pifont}
\usepackage{algorithm}
\usepackage{listings}
\usepackage{pythonhighlight}
\usepackage{float}
\usepackage{mathtools}
\usepackage{cuted}
\DeclareUnicodeCharacter{2212}{-}
\makeatletter
\newcommand\footnoteref[1]{\protected@xdef\@thefnmark{\ref{#1}}\@footnotemark}
\makeatother

\newcolumntype{P}[1]{>{\centering\arraybackslash}p{#1}}
\newlength\savewidth
\def\arrvline{\hfil\kern\arraycolsep\vline\kern-\arraycolsep\hfilneg}
\definecolor{mygray}{gray}{.9}
\definecolor{Highlight}{HTML}{39b54a}  

\definecolor{iblue}{rgb}{0.06, 0.75, 1.0}

\newcolumntype{x}[1]{>{\centering\arraybackslash}p{#1pt}}
\newcolumntype{z}[1]{>{\raggedright\arraybackslash}p{#1pt}}

\definecolor{citecolor}{HTML}{0071BC}
\definecolor{linkcolor}{HTML}{ED1C24}

\newcommand{\ourmodel}{{\fontfamily{ppl}\selectfont DiffTumor}}

\definecolor{cvprblue}{rgb}{0.21,0.49,0.74}
\usepackage[pagebackref,breaklinks,colorlinks,citecolor=cvprblue]{hyperref}

\def\paperID{1074} 
\def\confName{CVPR}
\def\confYear{2024}

\title{Towards Generalizable Tumor Synthesis}

\author{Qi Chen\textsuperscript{1} \quad Xiaoxi Chen\textsuperscript{2} \quad Haorui Song\textsuperscript{3} \quad Zhiwei Xiong\textsuperscript{1} \\ Alan~Yuille\textsuperscript{3} \quad Chen~Wei\textsuperscript{3,*} \quad Zongwei~Zhou\textsuperscript{3,}\thanks{Correspondence to Chen Wei (\href{mailto:weichen3012@gmail.com}{\textsc{weichen3012@gmail.com}}) and Zongwei Zhou (\href{mailto:zzhou82@jh.edu}{\textsc{zzhou82@jh.edu}})} \\[2.5mm]
\textsuperscript{1}University of Science and Technology of China\\
\textsuperscript{2}Shanghai Jiao Tong University \\
\textsuperscript{3}Johns Hopkins University \\[1.5mm]
{\small Code and Visual Turing Test:~\href{https://github.com/MrGiovanni/DiffTumor}{https://github.com/MrGiovanni/DiffTumor}}
}

\begin{document}

\maketitle

\begin{abstract}

Tumor synthesis enables the creation of artificial tumors in medical images, facilitating the training of AI models for tumor detection and segmentation. However, success in tumor synthesis hinges on creating visually realistic tumors that are \textbf{generalizable} across \ul{multiple organs} and, furthermore, the resulting AI models being capable of detecting real tumors in images sourced from \ul{different domains} (e.g., hospitals). This paper made a progressive stride toward generalizable tumor synthesis by leveraging a critical observation: early-stage tumors ($<$ 2cm) tend to have similar imaging characteristics in computed tomography (CT), whether they originate in the liver, pancreas, or kidneys. We have ascertained that generative AI models, e.g., Diffusion Models, can create realistic tumors generalized to a range of organs even when trained on a limited number of tumor examples from only one organ. Moreover, we have shown that AI models trained on these synthetic tumors can be generalized to detect and segment real tumors from CT volumes, encompassing a broad spectrum of patient demographics, imaging protocols, and healthcare facilities. 

\end{abstract}

\section{Introduction}
\label{sec:introduction}

\begin{figure*}[t]
\centerline{\includegraphics[width=1.0\textwidth]{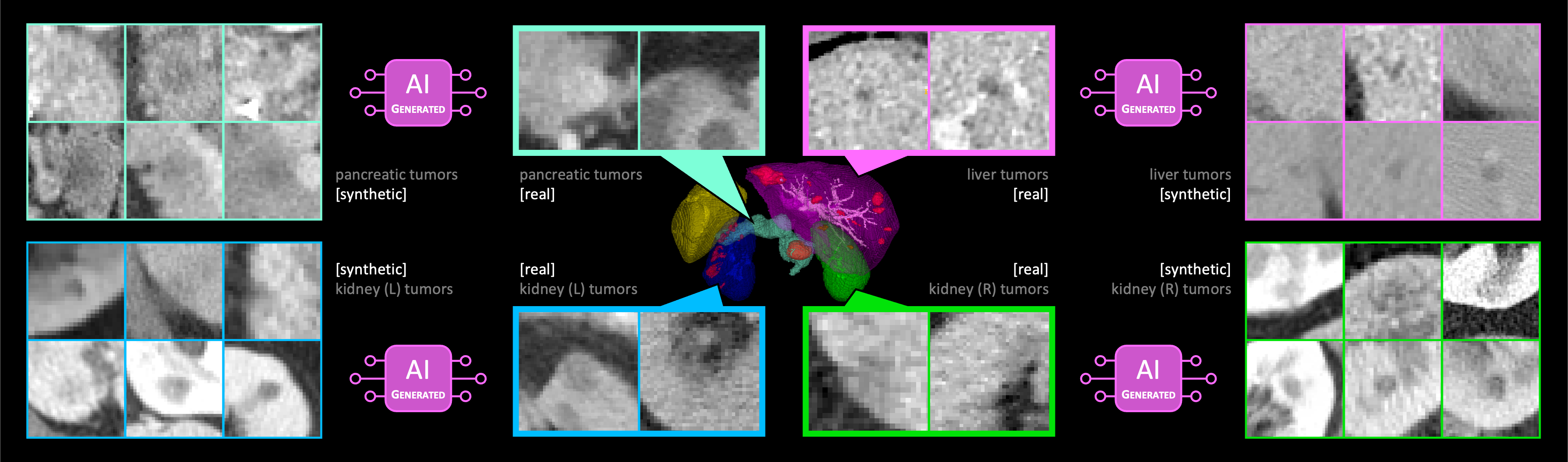}}
    \caption{
    \textbf{Generalizable tumor synthesis across organs.} Early-stage tumors present similar imaging characteristics in computed tomography (CT), whether they are located in the liver, pancreas, or kidneys. Leveraging this observation, we develop a generative AI model on a few examples of annotated tumors in a specific organ, e.g., the liver (in purple). This AI model (in purple), trained exclusively on liver tumors, can directly create synthetic tumors in those organs where CT volumes of annotated tumors are relatively scarce, e.g., the pancreas (in cyan) and kidneys (in blue and green). By integrating synthetic tumors into extensive CT volumes of healthy organs---routinely collected in clinical settings---we can substantially augment the training set for tumor segmentation. This enhancement can also significantly improve the AI generalizability across CT volumes sourced from diverse hospitals and patient demographics. 
    }
\label{fig:teaser}
\end{figure*}

Tumor synthesis enables the creation of artificial tumor examples in medical images~\cite{jordon2018pate,yoon2019time,chen2021synthetic}, it is particularly valuable when there is a dearth or complete absence of per-voxel annotated real tumors (e.g., early-stage tumors) for effective AI training. Typically, to train AI models for tumor detection in multiple ($N$) organs, annotated real tumor examples from each of these organs are necessary, and ideally, in substantial numbers~\cite{li2024well,kang2023label,liu2023clip,chou2024acquiring,zhu2022assembling}. Furthermore, AI models often fail to generalize across images from different hospitals, which may vary due to variations in imaging protocols, patient demographics, and scanner manufacturers~\cite{orbes2019multi,zhou2022interpreting,zhou2021towards}. The challenge amplifies with the need for extensive manual annotations, a task that could demand up to 25 human years for annotating just one tumor type~\cite{xia2022felix,chen2023cancerunit,abi2023automatic}. The task of collecting and annotating a comprehensive dataset encompassing tumors from multiple organs ($N$) and images from numerous hospitals ($M$) is daunting, considering both annotation cost and complexity ($N$\,$\times$\,$M$). We hypothesize that tumor synthesis could solve this challenge by creating various tumor types across non-tumor images from multiple hospitals, even when only one tumor type is available, thereby simplifying the complexity from $N$\,$\times$\,$M$ to $1$\,$\times$\,$M$.

Success in tumor synthesis hinges on creating visually realistic tumors that are \textit{generalizable} across multiple organs and, furthermore, the resulting AI models being \textit{generalizable} in detecting real tumors in images sourced from different hospitals. Previous studies have introduced generative models to create synthetic medical data (not limited to tumors) such as polyp detection from colonoscopy videos~\cite{shin2018abnormal}, COVID-19 detection from Chest X-ray~\cite{yao2021label,lyu2022pseudo,gao2023synthetic}, and diabetic lesion detection from retinal images~\cite{wang2022anomaly}---refer to \S\ref{sec:related_work} for a comprehensive review. However, these studies have primarily focused on enhancing the detection and segmentation of specific tumors without fully exploring the wider generalizability of these models across different organs and patient demographics.

This paper made a progressive stride toward generalizable tumor synthesis by leveraging a critical \textbf{observation}: \textit{early-stage tumors ($<$\,2cm) tend to have similar imaging characteristics in computed tomography (CT)}\footnote{Note that, owing to the public dataset constraints, we have only verified the similarity across early hepatocellular carcinoma and intrahepatic cholangiocarcinoma from the liver, pancreatic ductal adenocarcinoma from the pancreas, and renal cell carcinoma from kidneys.}. Early-stage tumors typically present small, round, or oval shapes with minimal deformation and exhibit relatively simple and uniform textures in CT volumes~\cite{choi2014ct}. Hence, early tumors in parenchymal organs (e.g., liver, spleen, pancreas, adrenal glands, and kidneys) should appear similarly, as shown in \figureautorefname~\ref{fig:teaser}. The major difference is the contrast between the tumors and background organs or other anatomical structures rather than the tumors themselves. Using four public datasets and our proprietary datasets, \S\ref{sec:hypothesis} verifies the similarity of early-stage tumors across various organs. 

Leveraging this observation, we introduce a novel framework, termed \ourmodel, that can learn the common imaging characteristics of tumors across various organs, and the generated synthetic tumors are useful for training AI models to detect and segment real tumors from CT volumes of varying patient demographics. The development of \ourmodel\ is composed of three stages. 
\ding{172} Training an \textit{Autoencoder Model}---consisting of an encoder and decoder---on 9,262 unlabeled three-dimensional CT volumes. The use of large, diverse datasets can enhance the model's ability to generalize across CT volumes of different patient demographics and reduce the need for annotated tumor volumes for training Diffusion Models in the subsequent stages. The proxy task is image reconstruction, which facilitates the model in learning comprehensive latent features. 
\ding{173} Training a \textit{Diffusion Model}---a specific type of generative models---using latent features and tumor masks as conditions. Once trained, this model can generate latent features necessary for reconstructing CT volumes with tumors based on arbitrary masks. 
\ding{174} Training a \textit{Segmentation Model} using synthetic tumors, which are reconstructed by the decoder, and their corresponding masks. With a large repository of healthy CT volumes, our \ourmodel\ framework can produce a vast array of synthetic tumors, varying in location, size, shape, texture, and intensity, therefore fostering high-performing AI models for tumor detection/segmentation.

The key contributions of this paper are two-fold. \textbf{Firstly}, we have verified with feature analysis, reader studies, and clinical knowledge that early-stage tumors ($<$\,2cm) manifest with similar imaging characteristics across various organs in CT volumes, establishing the foundation for the development of generalizable tumor synthesis. \textbf{Secondly}, we have developed a three-stage tumor synthesis framework, \ourmodel, that trains generative models with minimal annotations (\figureautorefname~\ref{fig:property_less_annotations}; \textit{one annotated CT volume}), creates synthetic tumors in real-time (\figureautorefname~\ref{fig:property_realtime}; \textit{100 ms/tumor}), and improves early-stage tumor detection (\figureautorefname~\ref{fig:property_early_tumors}; \textit{improved sensitivity up to +28.6\%}). 
In summary, compared with training AI on extensively annotated CT volumes of real tumors, our \ourmodel\ is generalizable from two critical perspectives.

\begin{enumerate}

    \item \ourmodel\ can create visually realistic tumors \textit{generalizable} to a range of organs even when the diffusion model was trained on a limited number of tumor examples from a specific organ (\S\ref{sec:generalizable_organs}; \textit{+10.7\% DSC}).

    \item \ourmodel\ can develop an AI model to detect and segment real tumors \textit{generalizable} to a variety of CT volumes of varied patient demographics, imaging protocols, and healthcare facilities (\S\ref{sec:generalizable_hospitals}; \textit{+9.1\% DSC}).
    
\end{enumerate}

\begin{figure*}[t]
	\centering
\includegraphics[width=\linewidth]{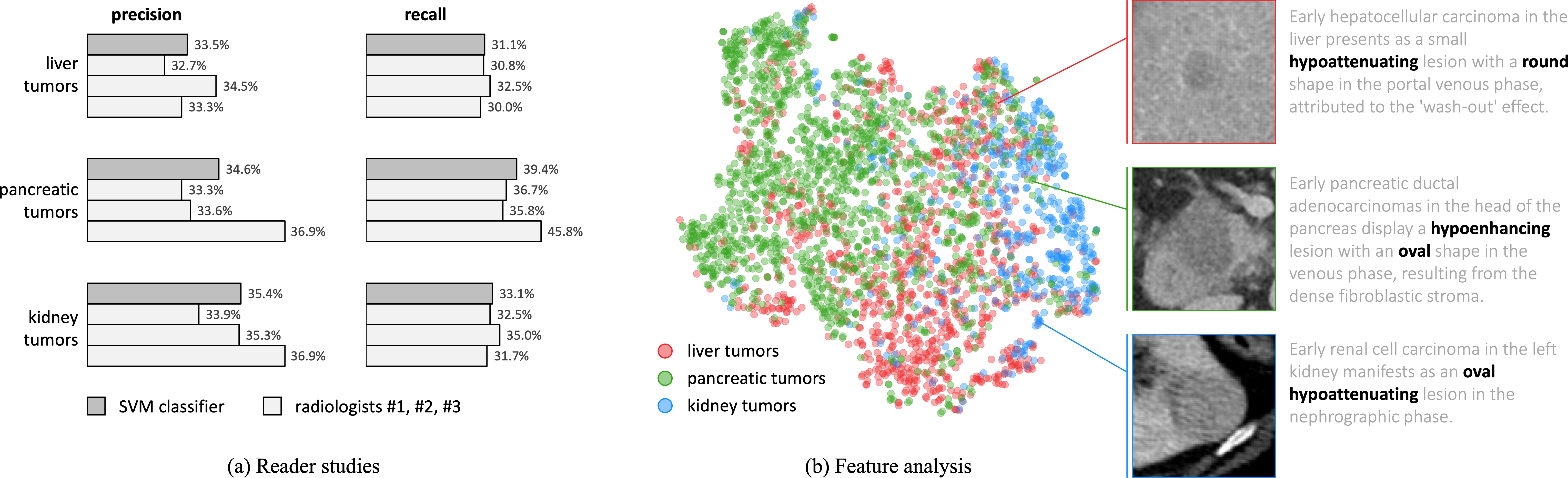}
	\caption{
    \textbf{Reader studies and feature analysis.} 
    We assess the performance of a support vector machine (SVM) classifier, using Radiomics features~\cite{chu2019utility}, and three expert radiologists in identifying the originating organs of cropped tumors. The SVM classifier is tasked with a three-way classification to ascertain whether a tumor originates from the liver, pancreas, or kidneys. In a similar test, radiologists examine the original CT images of these tumors. Reader study results on the \textit{left} panel indicate significant challenges for both the SVM classifier and the radiologists in accurately identifying the origin of early-stage tumors. The precision and recall scores for both methods closely resemble those of random guessing. Additionally, on the \textit{right} panel, we present a \textit{t}-SNE visualization of Radiomics features for tumors from the liver, pancreas, and kidneys. These results highlight the considerable similarity in features and images of early-stage tumors.
    }
	\label{fig:preliminary}
\end{figure*}

\section{Preliminary}
\label{sec:hypothesis}

We observe that \textit{early-stage tumors ($<$ 2cm)\footnote{Based on the TNM system, the most widely used staging system for classifying a malignancy tumor, we recognize a primary malignant tumor with a diameter less than 2 cm and no evidence of nearby lymph node involvement or metastasis as an early-stage tumor~\cite{burke2004outcome,ficarra2007tnm,rindi2012tnm}.} often share similar imaging characteristics in CT volumes, whether they originate in the liver, pancreas, or kidneys.} This observation, when confirmed, could have profound implications for generative AI in medical imaging. It suggests that generative AI might be trained on one tumor type, for which data and annotations are more easily obtained, and then applied to create various tumor types in different organs, where acquiring sufficient data can be challenging. Using synthetic tumors can substantially improve AI performance in tumor detection and segmentation in practice. In light of this, we have rigorously pursued the validation through four approaches as follows.

\smallskip\noindent\textbf{(1) Radiologist reader study.}
The objective of the reader study is to assess the ability of radiologists to recognize the organ class of early cancer. We uniformly crop 360 CT images of the tumor region from three abdominal organs, as per the annotations. In order to exclude the influence of surrounding organ textures on recognition, we only retain a small amount of organ textures in the tumor boundary region. Examples of CT crops used for the reader study are provided in \figureautorefname~\ref{fig:preliminary}, and more examples are in Appendix~\ref{sec:reader_study_examples_appendix}. Three expert radiologists, qualified under the Quality Standards Act, participate in the reader study. The recognition results are shown in \figureautorefname~\ref{fig:preliminary}(a). The nearly random probability of the precision and recall scores indicates that the appearance of early-stage tumors is so similar that even experienced radiologists have difficulty distinguishing the organ types of these tumors.

\smallskip\noindent\textbf{(2) Radiomics feature analysis.}
We now analyze the similarity in Radiomics features\footnote{We utilize the official Radiomics feature repository~\cite{van2017computational,wang2017comparison} to extract the appearance features, which include 3D shape-based features, gray level co-occurrence matrix, gray level run length matrix, gray level size zone matrix, neighboring gray-tone difference matrix, and gray level dependence matrix. More details can be seen in Appendix~\ref{sec:radiomics_feature_appendix}.} of early-stage tumors. \textit{Quantitatively}, we train three types of learning-based classifiers, including support vector machine (SVM), Random Forests, and  AdaBoost, to identify the organ types of early-stage tumors. To draw a general conclusion, we conducted ten repeated experiments and calculated the precision and recall scores of these classifiers in both the training and test sets. The final results for SVM show that the precision and recall for the training set are close to 1, indicating that SVM is well-trained and capable of learning a decent decision boundary for the training set. However, the precision scores for the test set are nearly equivalent to random probability, as shown in \figureautorefname~\ref{fig:preliminary}(a). Similarly, Random Forest achieves a precision of 50.3\% and a recall of 54.9\%, while  AdaBoost achieves a precision of 35.4\% and a recall of 49.5\%. This suggests that even a well-trained classifier struggles to recognize the organ types of unseen early-stage tumors. \textit{Qualitatively}, \figureautorefname~\ref{fig:preliminary}(b) visualizes the feature mapping in a two-dimensional space using \textit{t}-SNE. The appearance features of early-stage tumors are distributed in a joint feature space, and there is no separation for different organ types. 

\smallskip\noindent\textbf{(3) Deep feature analysis.}
We investigate the similarity of early tumors across different organs using deep features extracted by ResNet and DenseNet. These two networks are trained to classify the types of organs affected by early-stage tumors. ResNet achieves a precision of 59.7\% and a recall of 55.6\%; DenseNet achieves a precision of 44.3\% and a recall of 61.1\%. As seen, no matter whether using hand-craft features or deep features, the results reached a consistent observation---\textit{none of the classifiers can distinguish early tumors correctly among the three organs}.

\begin{figure*}[t]
	\centering
\includegraphics[width=1.0\linewidth]{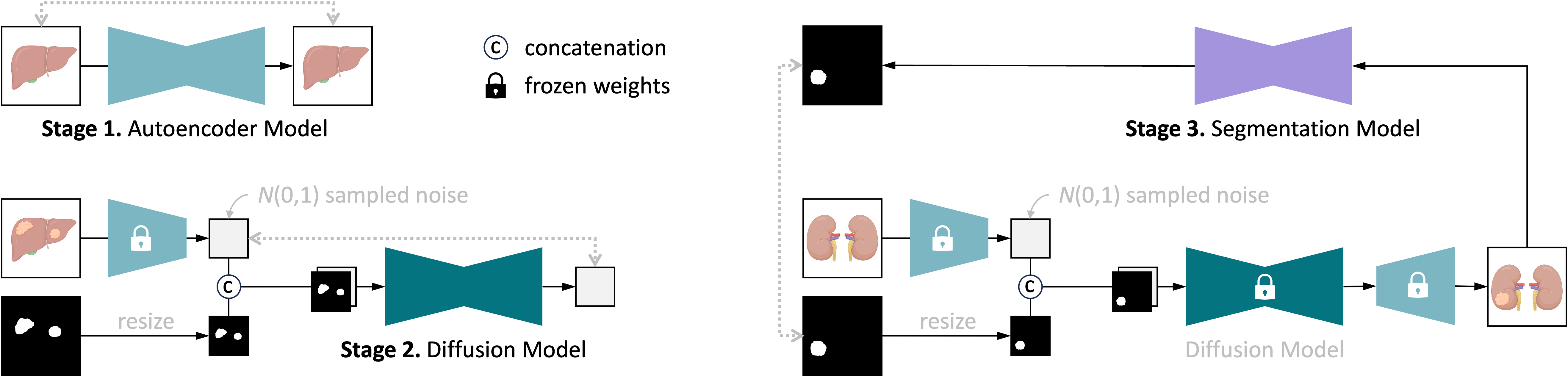}
	\caption{
    \textbf{Overview of the \ourmodel\ framework.} 
    Towards generalizable tumor synthesis, developing our \ourmodel\ involves three stages. \ding{172} Training an \textit{Antoencoder Model}---consisting of an encoder and decoder---to learn comprehensive latent features. The learning task here is image reconstruction performed on 9,262 unlabeled three-dimensional CT volumes. Both the trained encoder and decoder will be used in subsequent stages. 
    \ding{173} Training a \textit{Diffusion Model}---a specific type of generative models---using latent features and tumor masks as conditions. Once trained, this model can generate latent features necessary for reconstructing CT volumes with tumors based on arbitrary masks. 
    \ding{174} Training a \textit{Segmentation Model} using CT volumes of synthetic tumors, which are reconstructed by the decoder. With a large repository of healthy CT volumes, our \ourmodel\ framework can produce a vast array of synthetic tumors, varying in location, size, shape, texture, and intensity, therefore fostering high-performing AI models for tumor detection/segmentation.
    }
	\label{fig:method}
\end{figure*}

\smallskip\noindent\textbf{(4) Clinical evidence and justification.}
Tumorigenesis is a gradual, multi-step process involving cellular and histological changes that culminate in successively malignant lesions~\cite{choi2014ct}. Histologically, early-stage tumors often exhibit well-to-moderately differentiated neoplastic cells with mild atypia, limited hemorrhage, and necrosis~\cite{chu2017diagnosis, dunnick2016renal,ayuso2018diagnosis}.  This cellular similarity leads to shared imaging features across various parenchymal organs (e.g., liver, pancreas, spleen, adrenal gland, kidney).  Early-stage tumors typically appear as relatively homogeneous nodules with indistinct margins and small diameters in CT images~\cite{skarin2015atlas}. These consistent characteristics, observed across populations, ages, and genders, suggest that tumor synthesis models could learn and generalize shared imaging features across organs.

\begin{itemize}
    
    \item \textbf{Liver tumors:} 
    Hepatocellular carcinoma (HCC) in the early stage presents as a small, well-differentiated nodule with minimal metastatic potential~\cite{fowler2021pathologic}. Active neoangiogenesis leads to reduced portal triads and an isoattenuating or hypoattenuating appearance (wash-out) in the venous phase compared to surrounding parenchyma~\cite{ayuso2018diagnosis}.

    \item \textbf{Pancreatic tumors:} Multiphase CT with intravenous contrast is preferred for diagnosing suspected pancreatic lesions~\cite{laeseke2015combining}. Most pancreatic ductal adenocarcinomas (PDACs) exhibit hypoenhancement relative to surrounding tissue due to their dense fibroblastic stroma~\cite{elbanna2020imaging}.

    \item \textbf{Kidney tumors:} 
    CT is the gold standard for evaluating renal cell carcinoma (RCC)~\cite{leveridge2010imaging}. Clear cell RCC, the most common subtype, typically presents as a small, hypoattenuating renal lesion with surrounding homogenous enhancement in the nephrographic phase~\cite{pmid29668296}.
\end{itemize}

\section{\ourmodel}
\label{sec:method}

\subsection{Autoencoder Model}
\label{subsec:Autoencoder_Model}

Diffusion models directly applied to three-dimensional CT volumes incur significant computational costs. To address this, Latent Diffusion Models (LDMs)~\cite{rombach2022high} operate within a compressed, lower-dimensional latent space. Following this approach, we construct our diffusion model within the latent space of 3D CT volumes. Our first step involves training a 3D autoencoder to learn meaningful, compressed latent representations. We adapt the Vector Quantized Generative Adversarial Networks (VQGAN)~\cite{esser2021taming} architecture, replacing 2D convolutions with their 3D counterparts.

Formally, we denote a CT sub-volume as $\boldsymbol{x}$\,$\in$\,$\mathbb{R}^{H \times W \times D}$, where $H$ denotes the height, $W$ the width, and $D$  the depth. The CT sub-volume $\boldsymbol{x}$ is first converted to latent features $\boldsymbol{z}$ by an encoder $f$ and a quantization operation $\mathbf{q}$, \ie, $\boldsymbol{z}$\,$=$\,$\mathbf{q}\left(f(\boldsymbol{x})\right)$\,$\in$\,$\mathbb{Z}^{h \times w \times d}$, where $h$ denotes the feature height, $w$ the feature width, and $d$ the feature depth. In the vector quantization step, the latent features $\boldsymbol{z}$ are quantized into $\boldsymbol{c}_z$\,$\in$\,$\mathbb{R}^{h \times w \times d \times c}$ by replacing each one with its closest corresponding codebook vector in the learned codebook $\mathcal{C}=\left\{\boldsymbol{c}_i\right\}_{i=1}^K$. $K$ is the codebook size. Finally, a decoder $g$ reconstructs the latent from $\boldsymbol{c}_z$ to $\hat{\boldsymbol{x}}=g\left(\boldsymbol{c}_z\right)$. The loss is a summation of three terms:
\begin{equation}
\footnotesize
    \mathcal{L}_{\text {recon}} + \mathcal{L}_{\text {codebook}} + \alpha\mathcal{L}_{\text {commit}},
\end{equation}
where $\mathcal{L}_{\text {recon}}$\,$=$\,${\|\boldsymbol{x}-\hat{\boldsymbol{x}}\|_1}$, $\mathcal{L}_{\text {codebook}}$\,$=$\,$ {\left\|\operatorname{sg}\left[f(\boldsymbol{x})\right]-\boldsymbol{c}_z\right\|_2^2}$, and $\mathcal{L}_{\text {commit}}$\,$=$\,${\alpha\left\|\operatorname{sg}\left[\boldsymbol{c}_z\right]-f(\boldsymbol{x})\right\|_2^2}$. $\operatorname{sg}[\cdot]$ denotes the stop-gradient operation and $\alpha$ the coefficient.

In addition to these three loss terms, we also adopt a perceptual loss and a discriminator to improve the reconstruction quality. For 3D CT reconstruction, we adopt a 3D volume discriminator $D_v$ to penalize implausible artifacts for the
3D reconstruction of $\hat{\boldsymbol{x}}$, and a 2D slice discriminator $D_s$ to encourage per-slice quality. To stabilize the GAN training, we add the feature matching losses $\mathcal{L}_{\text {match }}$.
Moreover, due to the CT volumes being preprocessed to isotropic volume, we constrain the high-frequency texture for all three planes of $\hat{\boldsymbol{x}}$ by using perceptual loss for projected reconstruction slices $\hat{\boldsymbol{x}}_{HW}$\,$\in$\,$\mathbb{R}^{H \times W}$, $\hat{\boldsymbol{x}}_{HD}$\,$\in$\,$\mathbb{R}^{H \times D}$, $\hat{\boldsymbol{x}}_{WD}$\,$\in$\,$\mathbb{R}^{W \times D}$.
The overall objective of the Autoencoder is:
\begin{equation}
\footnotesize
    \begin{aligned}
        & \min _{f, g, \mathcal{C}}\left( \mathcal{L}_{\text {recon }}+\mathcal{L}_{\text {codebook }}+\alpha \mathcal{L}_{\text {commit }}+\mathcal{L}_{\text {match }}+ \mathcal{L}_{\text {perceptual }}\right) \\
        & + \min _{f, g, \mathcal{C}}\left(\max _{D_s, D_v}\left(\mathcal{L}_{\text {disc }}\right)\right),
\end{aligned}
\end{equation}
where
\begin{equation}
\footnotesize
\mathcal{L}_{\mathrm{disc}}=\log D_{s / v}(\boldsymbol{x})+\log \left(1-D_{s / v}(\hat{\boldsymbol{x}})\right), 
\end{equation}
\begin{equation}
\footnotesize
\mathcal{L}_{\text {match }}=\sum_i \left\|D_{s / v}^{(i)}(\hat{\boldsymbol{x}})-D_{s / v}^{(i)}(\boldsymbol{x})\right\|_1.
\end{equation}
$D_{s / v}^{(i)}$ denotes the $i_{th}$ layer of discriminators.

\subsection{Diffusion Model}
\label{subsec:Diffusion_Model}

We aim to synthesize realistic and diverse CT volumes with tumors to facilitate the training of the tumor segmentation model.
Given the fact that healthy CT volumes are much more accessible than CT volumes with tumors, we focus only on tumor synthesis, and we do \textit{not} intend to model organ textures outside of the tumors, which can be easily obtained from healthy CT volumes.
To be specific, our diffusion model is conditioned on both a tumor mask that indicates the shape and location of tumors in the latent feature and the healthy region of CT volumes. 

Formally, given a pair of tumor-present CT volume  $\boldsymbol{x}_0$ and the mask of its tumor region $\boldsymbol{m}$, the diffusion model is conditioned on both the tumor mask $\boldsymbol{m}$ and the healthy region $\boldsymbol{z}_0^{\textrm{healthy}}$\,$\coloneqq$\,$(1-\boldsymbol{m}) \odot \boldsymbol{z}_0$. The diffusion model approximates the distribution of the latent features of tumor-present CT volumes.
In the forward process, the latent feature $\boldsymbol{z}_0$ is gradually converted to white Gaussian noise $\boldsymbol{z}_T \sim \mathcal{N}(0,1)$ by recursively adding a small amount of Gaussian noise $T$ times following the Markov process below:
\begin{equation}
\footnotesize
    p\left(\boldsymbol{z}_t \mid \boldsymbol{z}_{t-1}\right)=\mathcal{N}\left(\boldsymbol{z}_t ; \sqrt{1-\beta_t} \boldsymbol{z}_{t-1}, \beta_t \mathbf{I}\right),
\end{equation}
where $t \in [1,2,..,T]$ denotes the timestep and $\beta_{1:T}$ is the variance schedule of noise.

In the inference, we synthesize the latent feature of CT volumes by sampling from $p\left(\boldsymbol{z}_0 \mid  \boldsymbol{z}_0^{\textrm{healthy}},\boldsymbol{m}\right)$, which is approximated by recursively sampling
from $p\left(\boldsymbol{z}_{t-1},\mid \boldsymbol{z}_t,\boldsymbol{z}_0^{\textrm{healthy}}, \boldsymbol{m}\right)$. The training objective of our diffusion model~\cite{ho2020denoising} is as follows:
\begin{equation}
\footnotesize
\mathbb{E}_{\boldsymbol{z}_0, \epsilon \sim \mathcal{N}(0,1), t}\left[\left\|\epsilon-\epsilon_\theta\left(\boldsymbol{z}_t, \boldsymbol{z}_0^{\textrm{healthy}}, \boldsymbol{m},t\right)\right\|_2^2\right],
\end{equation}
where $\epsilon_\theta(\cdot,t)$ is a 3D U-Net with interleaved self-attention layers and convolutional layers~\cite{ho2020denoising,nichol2021improved} that predicts the noise given the input. To reduce the heavy computational cost for 3D CT volumes, we factorize the self-attention over the entire 3D data to first only attend to each 2D slide and then attend to the depth dimension, inspired by 3D video Transformers~\cite{bertasius2021space,arnab2021vivit}. This design largely reduces the computation cost of the self-attention layers in the 3D U-Net.

\subsection{Segmentation Model}
\label{subsec:Segmentation_Model}

We construct a large-scale database of healthy CT volumes as a basis for our tumor synthesis method. This database includes 1,246 volumes with healthy livers, 1,901 with healthy pancreases, and 1,005 with healthy kidneys, ensuring diversity across ages, genders, nationalities, and acquisition protocols. Following Hu~\etal~\cite{hu2023label}, we generate realistic tumor-like shapes using ellipsoids and refine them with expert radiologist feedback for clinical plausibility (implementation details in Appendix~\ref{sec:dataset_implementation_details_appendix}). By combining these generated tumor masks with the healthy CT volumes (Figure~\ref{fig:method}--Stage 3), we synthesize tumors across various domains, promoting generalizability in our model.

\section{Experiments \& Results}
\label{sec:result}
\noindent\textbf{Real-tumor datasets.} LiTS~\citep{bilic2019liver}, MSD-Pancreas~\cite{antonelli2021medical}, and KiTS~\cite{heller2020international} were used for training and testing Segmentation Models on the liver, pancreas, and kidneys, respectively. We performed 5-fold cross-validation on 118 tumor CT volumes for LiTS and 120 tumor CT volumes for MSD-Pancreas and KiTS.

\smallskip\noindent\textbf{Healthy CT datasets.} We collect a large repository of healthy CT volumes. Due to the computational cost and memory limitation for training, we only randomly selected 120 healthy CT volumes for the kidney and pancreas, respectively. For liver, we adopt the same healthy CT volumes as Hu~\etal~\cite{hu2023label}. More details about dataset and implementation can be found in Appendix~\ref{sec:dataset_implementation_details_appendix}.

\subsection{Visual Turing Test}
\label{sec:visual_turing_test}

We conduct the Visual Turing Test on 240 CT volumes for three organs, respectively, where 120 volumes are with real tumors, and the remaining 120 volumes are with synthesized tumors by our method. Four radiologists, with varying levels of experience ranging from junior to senior and professional, are involved in this test. The total Visual Turing Test took 144 hours (2,880 CTs). Following Hu~\etal~\cite{hu2023label}, each sample is inspected in a 3D view to be classified as either real or synthetic, allowing for the observation of a continuous slice sequence. The testing results are shown in Table~\ref{tab:reader_studies}. All radiologists are able to identify real tumors with a high sensitivity score (above 90\%). This indicates their familiarity with the characteristics of real tumors. However, the low specificity scores (below 40\%) on the three types of tumors for radiologists R1 and R3 suggest that the synthetic data strongly resembles real tumors, leading to most synthetic tumors being misidentified as real ones. As for R2 and R4, who have more experience, the specificity scores are higher than those of R1 and R3, approximating 50\%. This indicates that nearly 50\% of synthetic samples are still incorrectly identified as real tumors. These results confirm the efficacy of \ourmodel\ in generating visually realistic tumors.

\begin{table}[t]
    \centering
    \scriptsize
    \begin{tabular}{p{0.11\linewidth}p{0.18\linewidth}P{0.1\linewidth}P{0.1\linewidth}P{0.1\linewidth}P{0.1\linewidth}}
    \toprule
     & & R1 & R2 & R3 & R4 \\
    \midrule
    \multirow{3}{*}{\makecell[l]{liver}} & sensitivity (\%) &98.3 &99.2 &100 &100 \\
     & specificity (\%) &31.7 &53.3 &39.2 &45.8 \\
     & accuracy (\%) &65.0  &76.3 &69.6 &72.9 \\
    \midrule
    \multirow{3}{*}{\makecell[l]{pancreas}} & sensitivity (\%) &96.7 &100 &100 &98.3\\
     & specificity (\%) &22.5 &44.2 &34.2 &38.8 \\
     & accuracy (\%) &59.6  &72.1 &67.1 & 68.3\\
    \midrule
     \multirow{3}{*}{\makecell[l]{kidney}} & sensitivity (\%) &95.8 &98.3 &99.2 &97.5 \\
     & specificity (\%) &36.7  &55.0 &40.8 &51.7 \\
     & accuracy (\%) &66.3  &76.7 &70.0 &74.6 \\
    \bottomrule
    \end{tabular}
    \begin{tablenotes}
        \item positives: real tumors ($N$ = 120); negatives: synthetic tumors ($N$ = 120).
    \end{tablenotes}
    
    \caption{
    \textbf{Visual Turing Test} over three organs has been conducted with four radiologists (R1--R4). Each radiologist was provided with 240 three-dimensional CT volumes of each organ, including 120 scans with real tumors and the remaining 120 with synthetic ones. Radiologists were tasked to label each CT volume as \textit{real} or \textit{synthetic}. A lower specificity score indicates a higher number of synthetic tumors being identified as real.
    }
    \label{tab:reader_studies}
\end{table}

\subsection{Generalizable to Different Organs}
\label{sec:generalizable_organs}

DiffTumor can generate visually realistic tumors generalizable to a range of organs, although Diffusion Model is only trained on a specific organ tumor.
In order to verify the effectiveness of our method's generalization capacity across different organs, we conducted comparative experiments across three different abdominal organs. This involved training all Segmentation Models on tumor data from a single organ, and then applying that training to the other two organs. For the results of our method, we train \ourmodel\ on source organ data, then utilize healthy CT volumes to synthesize tumors in the target organ, which are used for further training of Segmentation Models. To showcase the broad applicability of our synthetic data, we compare the early-stage tumor detection capabilities across three commonly used backbones. The generalization result, shown in \tableautorefname~\ref{tab:generalizable_organs}, suggests that it is difficult for Segmentation Models trained on real data to generalize across different organs, leading to poor performance in early-stage tumor detection. Hu~\etal~\cite{hu2023label} introduces a modeling-based method, which can maintain consistent sensitivity scores for the same target, regardless of the source domain.
The generalization ability of \ourmodel\ across organs surpasses that of most models, except in the setting that generalizing tumors from kidney to liver. Moreover, we demonstrate the strength of \ourmodel\ used as an augmentation method for real tumors in the same organ, as shown in \tableautorefname~\ref{tab:generalizable_organs}. In particular, there is a notable improvement of 10.7\% in the Dice Similarity Coefficient (DSC) for kidney tumors when using nnU-Net backbone. Furthermore, a decrease in the standard deviations of the DSC scores suggests that Segmentation Models become more stable. The significant improvement in DSC scores across all three organs proves that \ourmodel\ is an effective data augmentation method to enhance the performance of Segmentation Model.

\begin{table}[t]
    \centering
    \scriptsize
    \begin{tabular}{p{0.14\linewidth}p{0.16\linewidth}P{0.15\linewidth}P{0.15\linewidth}P{0.15\linewidth}}
    \multicolumn{5}{l}{\textit{Early-stage tumor detection performance (tumor-wise Sensitivity \%).}} \\
    \toprule
    \multicolumn{2}{l}{source $\backslash$ target} & liver & pancreas & kidneys \\
    \midrule
    \multirow{3}{*}{\makecell[l]{liver}} & real tumors & 75.6 &0 & 2.4\\
     & Hu~\etal~\cite{hu2023label} &77.8 &56.3 &52.4\\
     & \ourmodel &\cellcolor{iblue!10}\textbf{82.2} &\cellcolor{iblue!10}\textbf{56.3}&\cellcolor{iblue!10}\textbf{76.2}\\
    \midrule
    \multirow{3}{*}{\makecell[l]{pancreas}} & real tumors &0.7  &  64.3 &0 \\
     & Hu~\etal~\cite{hu2023label} &74.1 &67.0 &52.4\\
     & \ourmodel &\cellcolor{iblue!10}\textbf{75.3}&\cellcolor{iblue!10}\textbf{71.4} &\cellcolor{iblue!10}\textbf{71.4}\\
    \midrule
    \multirow{3}{*}{\makecell[l]{kidney}} & real tumors &0.1  & 0 &  50.0 \\
     & Hu~\etal~\cite{hu2023label} &\cellcolor{iblue!10}\textbf{74.1} &56.3 &66.7\\
     & \ourmodel &68.8 &\cellcolor{iblue!10}\textbf{61.6} &\cellcolor{iblue!10}\textbf{78.6} \\
    \bottomrule
    \end{tabular}
    \begin{tabular}{p{0.14\linewidth}p{0.16\linewidth}P{0.15\linewidth}P{0.15\linewidth}P{0.15\linewidth}} \\
    \multicolumn{5}{l}{\textit{All-stage tumor segmentation performance (DSC \%).}} \\
    \toprule
    backbone & method & liver & pancreas & kidneys \\
    \midrule
    \multirow{2}{*}{\makecell[l]{U-Net}} & real tumors &62.3$\pm$28.3  &  56.0$\pm$24.8 & 75.1$\pm$27.2 \\
     & \ourmodel & \cellcolor{iblue!10}\textbf{70.9$\pm $21.1}& \cellcolor{iblue!10}\textbf{64.8$\pm$24.5} & \cellcolor{iblue!10}\textbf{84.2$\pm$9.5}\\
    \midrule
    \multirow{2}{*}{\makecell[l]{nnU-Net}} & real tumors &64.3$\pm$26.5  & 59.9$\pm$23.8 & 73.8$\pm$20.9 \\
     & \ourmodel & \cellcolor{iblue!10}\textbf{73.6$\pm$18.1} & \cellcolor{iblue!10}\textbf{63.6$\pm$27.7} & \cellcolor{iblue!10}\textbf{84.5$\pm$11.5} \\
    \midrule
    \multirow{2}{*}{\makecell[l]{SwinUNETR}} & real tumors & 65.1$\pm$23.5 &52.2$\pm$31.2 & 80.6$\pm$19.6\\
     & \ourmodel & \cellcolor{iblue!10}\textbf{71.4$\pm$19.1} & \cellcolor{iblue!10}\textbf{62.2$\pm$26.1} & \cellcolor{iblue!10}\textbf{85.1$\pm$8.7} \\
    \bottomrule
    \end{tabular}
    \caption{
    \textbf{Generalizable to different organs:} comparison of generalization for early-stage tumor detection under different source organs. The scores in bold represent the best performance in each domain. \ourmodel\ achieves the best performance in almost all domains.  Furthermore, \ourmodel\ serves as an effective data augmentation method for real tumors in three abdominal organs, yielding substantial improvements in all-stage tumor segmentation. Additional results for different segmentation backbones with more metrics can be found in the Appendix~\ref{sec:generalizable_organs_appendix}. 
    }
    \label{tab:generalizable_organs}
\end{table}

\subsection{Generalizable to Different Demographics}
\label{sec:generalizable_hospitals}
\begin{figure}[t]
	\centering
	\includegraphics[width=\linewidth]{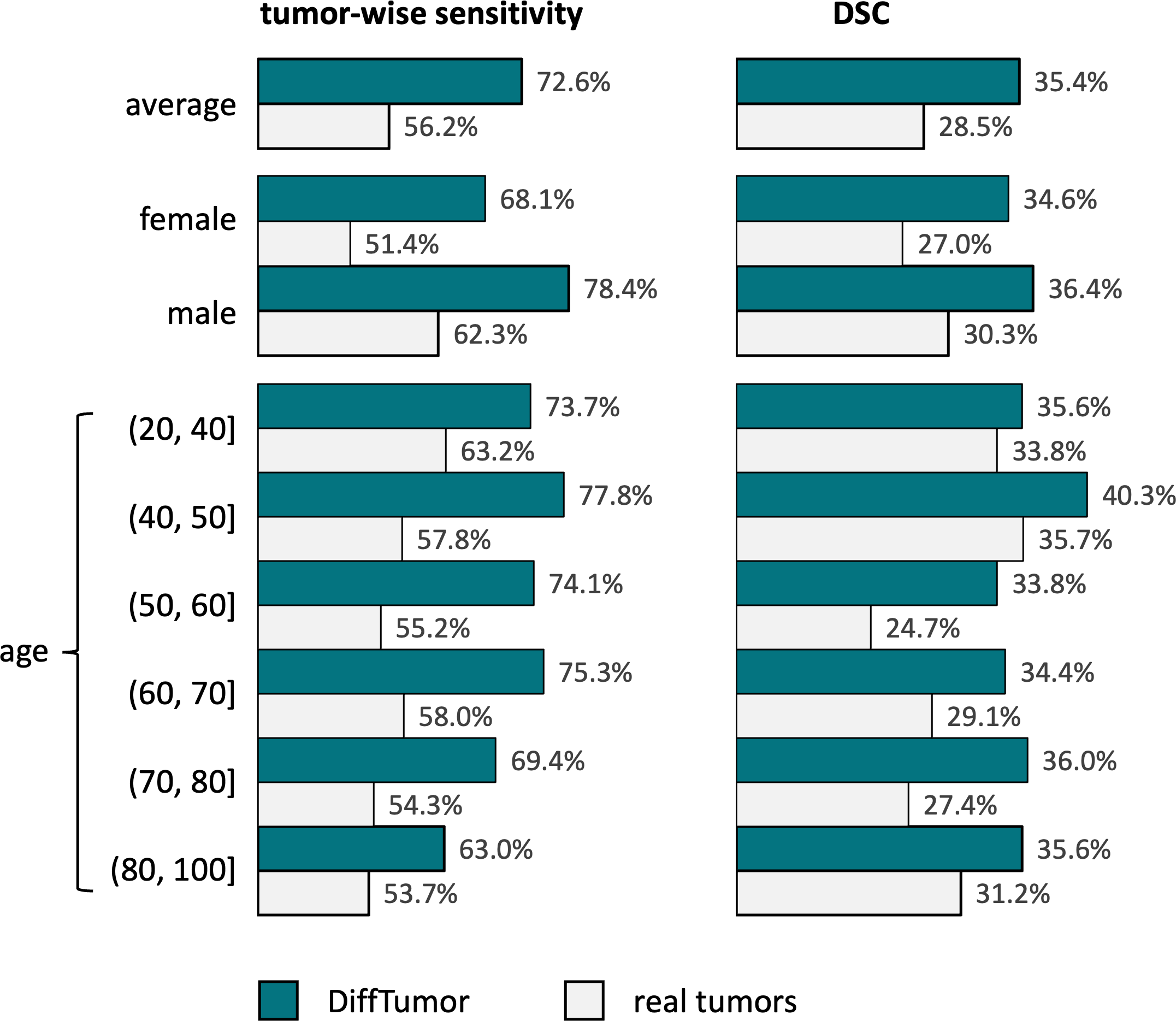}
	\caption{\textbf{Generalizable to various demographics.} Tumor detection and segmentation enhancement for individuals across various age groups and genders. \ourmodel\ can consistently boost tumor detection and segmentation performance by a significant margin in each patient group. Results of more segmentation backbones (\eg, nnU-Net and Swin UNETR) can be found in Appendix~\ref{sec:generalizable_hospitals_appendix}.
    }
	\label{fig:generalizable_hospitals}
\end{figure}

The ability of Segmentation Model to be generalizable to different demographics is critically important. It indicates that the model can effectively process CT scans from a diverse population, including various ages, genders, and ethnicities. To affirm the enhancement of \ourmodel\ for Segmentation Model to detect and segment real tumors across different individuals, we evaluate the generalization ability of Segmentation Model using a proprietary dataset at Hopkins~\cite{xia2022felix}. This dataset includes various real pancreatic tumors (PDAC and Cyst) from diverse patient demographics. We utilize \ourmodel\ with Diffusion Model trained on MSD-Pancreas to enhance Segmentation Model. More details about the dataset and experiment setting can be found in Appendix~\ref{sec:generalizable_hospitals_appendix}. 
\figureautorefname~\ref{fig:generalizable_hospitals} shows that our synthetic data can yield an average improvement of 6.9\% in DSC and 16.4\% in sensitivity with the U-Net backbone. In particular, the improvement for people aged 50--60 is significant, with an enhancement of 18.9\% in sensitivity and 9.1\% in DSC. For both males and females, there are noticeable performance improvements for tumor detection and segmentation. These results demonstrate that our synthetic data can provide valuable assistance in clinical tumor analysis for individuals across various age groups and genders.

\subsection{Advantages of DiffTumor}
\label{sec:property}

\begin{figure}[t]
	\centering
	\includegraphics[width=\columnwidth]{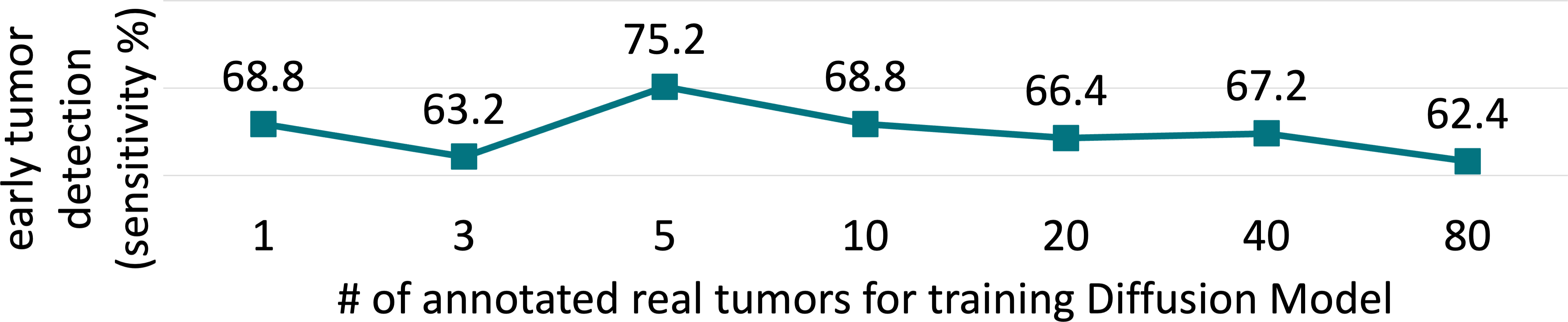}
	\caption{
    \textbf{Reduced annotations for Diffusion Model.}
    Diffusion Model, trained on annotated tumors in Stage \ding{173}, can generate synthetic tumors for the subsequent training of Segmentation Model in Stage \ding{174}. We investigate the relationship between the number of annotated real tumors required for Diffusion Model and the resultant performance of Segmentation Model. Results with varying numbers of annotated tumors reveal a surprising finding: extensive annotations are not necessary for tumor synthesis, contrary to the experience in computer vision~\cite{saharia2022palette,rombach2022high}. Notably, training Diffusion Model with only one annotated tumor seems to be sufficient. This efficiency is connected to our earlier observation in \S\ref{sec:hypothesis} that tumors, particularly in their early stages, tend to present similar appearances across different organs, thus facilitating the learning process of Diffusion Model with fewer annotated examples.
    }
	\label{fig:property_less_annotations}
\end{figure}

\noindent\textbf{(1) Reduced annotations for Diffusion Model.}
The quality of synthetic data produced by a generative model is typically heavily reliant on the quantity and diversity of the paired training data used during the training phase~\cite{chlap2021review,jaipuria2020deflating,ramesh2022hierarchical}.We study the relationship between the number of annotated real tumors needed for the Diffusion Model and the performance of the Segmentation Model. We find that the relationship between the amount of paired training data and the quality of synthetic data isn't always linear, as shown in \figureautorefname~\ref{fig:property_less_annotations}. In particular, \ourmodel\ only requires just one annotated tumor to train the Diffusion Model and generate synthetic tumors for the subsequent training of Segmentation Model. This contradicts the typical experience in computer vision~\cite{ramesh2022hierarchical}, which generally requires large-scale data for training. The results indicate that for training the Diffusion Model, particularly for early tumors, we can rely on a smaller number of real tumors. This finding could have important implications for the efficiency and cost-effectiveness of training \ourmodel.

\begin{figure}[t]
	\centering
	\includegraphics[width=\columnwidth]{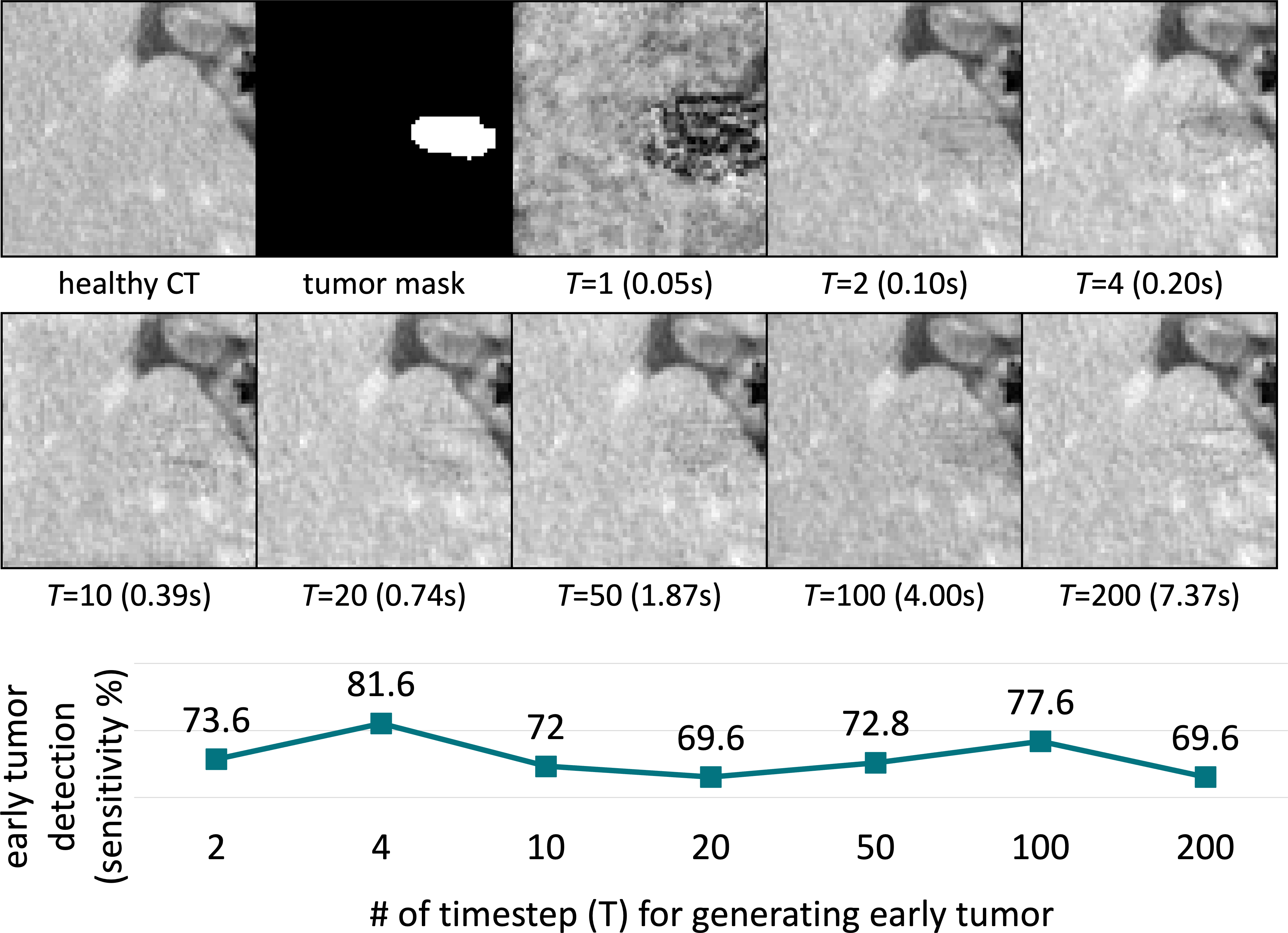}
	\caption{\textbf{Accelerated tumor synthesis.} The speed of generating synthetic tumors in Diffusion Model is significantly influenced by the timestep hyper-parameter (see figure). A faster tumor synthesis is preferred in Stage \ding{174} when training Segmentation Model, so we examine the impact of timestep on the performance of Segmentation Model. Our findings indicate that a timestep of 4, generating a tumor in 0.2 seconds, provides the most favorable results among various tested timesteps. Considering the trade-off between performance and efficiency, we chose a timestep of 4 for this study.
    }
	\label{fig:property_realtime}
\end{figure}

\smallskip\noindent\textbf{(2) Accelerated tumor synthesis.}
\label{sec:property_realtime}
The speed of tumor synthesis plays a crucial role in the practical application of synthetic data. Real-time synthes can significantly speed up the training process of Segmentation Model. The speed of generating synthetic tumors in Diffusion Model is significantly influenced by
the timestep $T$. We examine the impact of the timestep on the performance of Segmentation Model. The synthetic quality using DDPM sampling~\cite{ho2020denoising} with different timestep is illustrated in ~\figureautorefname~\ref{fig:property_realtime}. As can be seen, when $T=1$, the model collapses and fails to synthesize realistic textures for both the organ and tumor textures. Consequently, using these synthetic data to train the Segmentation Model results in poor performance. However, when $T$ is increased to more than 1, the corresponding texture can be well-generated, leading to good performance in the Segmentation Model. In consideration of the trade-off between performance and efficiency, we default to a timestep of $T=4$ for early tumor synthesis. This balance allows for the generation of high-quality synthetic data while maintaining a reasonable efficiency level.

\begin{figure}[t]
	\centering
	\includegraphics[width=\columnwidth]{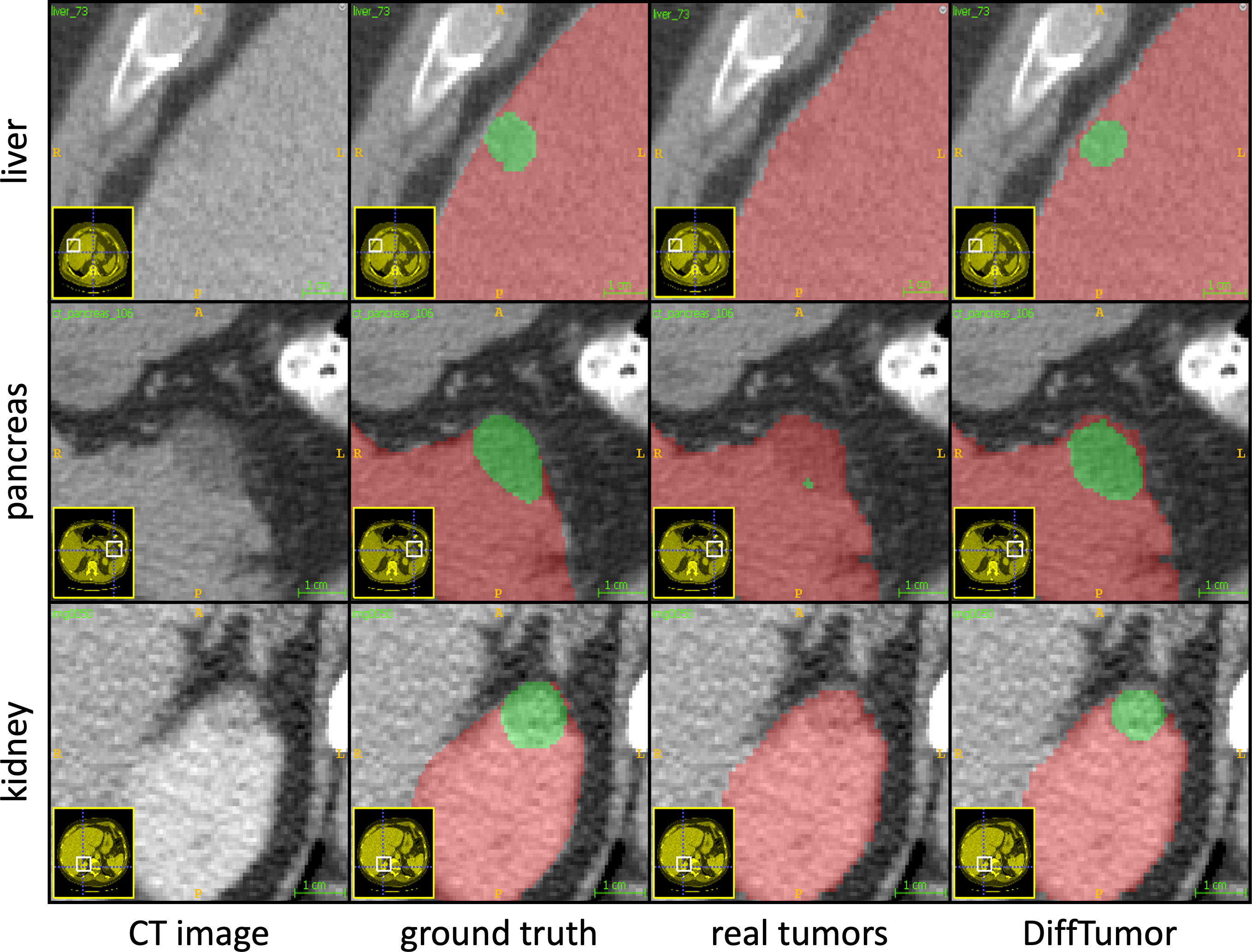}
	\caption{
    \textbf{Enhanced early tumor detection.} We analyzed failure cases of AI models trained on real tumors, identifying that these models often overlook early-stage tumors characterized by blurry boundaries, small sizes, and peripheral organ locations. Our \ourmodel\ enhances the detection and segmentation of these challenging tumors by extensively generating small tumors for AI training (evidenced in \tableautorefname~\ref{tab:generalizable_organs}). Further visualizations and examples are available in Appendix~\ref{sec:generalizable_organs_appendix}.
    } 
	\label{fig:property_early_tumors}
\end{figure}

\smallskip\noindent\textbf{(3) Improved early tumor detection.}
\label{sec:early_tumors}
Detecting tumors in their early stages can greatly increase the chances of successful treatment and survival. However, obtaining early-stage cancer data is challenging in practice and such cases in real datasets remain scarce. This  limits the AI model's ability to detect early tumors. As shown in \figureautorefname~\ref{fig:property_early_tumors}, there are several failure cases for Segmentation Model trained on real data. However, with the incorporation of our synthetic data, Segmentation Models' capability to detect early-stage tumors improves significantly. This is one of the primary reasons why \ourmodel\ can achieve the best performance as displayed in Table~\ref{tab:generalizable_organs}. This demonstrates the value and efficacy of synthetic data in enhancing early tumor detection.

\section{Related Work}
\label{sec:related_work}

\smallskip\noindent\textbf{Generative models} such as Energy-Based Models~\cite{lecun2006tutorial,zhao2016energy, du2019implicit}, Variational Autoencoders (VAE)~\cite{kingma2013auto,kingma2014semi,kingma2019introduction}, Generative Adversarial Networks (GAN)~\cite{goodfellow2014generative,goodfellow2016deep, goodfellow2020generative, creswell2018generative,jordon2018pate,yoon2019time,chen2022mask}, and normalizing flows~\cite{papamakarios2021normalizing, kobyzev2020normalizing,yu2021fastflow} have shown significant potential in creating realistic images.
Among these, Diffusion Models~\cite{sohl2015deep,ho2020denoising,song2020score} and their variants~\cite{vahdat2021score,kim2021maximum,rombach2022high} have recently emerged as particularly advanced in image generation.
In the medical field, generative models have been effectively utilized for tasks like image-to-image translation~\cite{lyu2022conversion,meng2022novel,ozbey2023unsupervised}, reconstruction~\cite{song2021solving,xie2022measurement}, segmentation~\cite{fernandez2022can,kim2022diffusion,wolleb2022diffusion}, image denoising~\cite{gong2023pet}, and anomaly detection~\cite{wyatt2022anoddpm,siddiquee2019learning,xiang2023squid}. In this work, we focus on generating tumors in abdominal organs based on the textures of the surrounding organs, which significantly reduces the annotated data required for training. Refer to Appendix~\ref{sec:discussion_appendix} for a more comprehensive comparison with our \ourmodel.

\smallskip\noindent\textbf{Tumor synthesis} that is widely effective for a variety of organs is an attractive topic.
Successful works related to tumor synthesis based on various medical modalities include colon polyp synthesis in colonoscopy videos~\cite{shin2018abnormal}, tumor cell synthesis in fluorescence microscopy images~\cite{horvath2022metgan}, synthesized brain tumors~\cite{billot2023synthseg,zhang2023self} and myocardial pathology~\cite{zhang2024lefusion} in MRI, lung nodule synthesis in CT images~\cite{han2019synthesizing,yang2019class,jin2021free}, and lesion in dermatoscopic images~\cite{du2023boosting}. Additionally, there are many works on synthesizing non-cancerous lesions such as COVID-19 lesion synthesis in chest CT~\cite{lyu2022pseudo,yao2021label}, and diabetic lesion synthesis in retinal images~\cite{wang2022anomaly}.
Recent studies have improved the realism of synthetic tumors in the liver~\cite{zhang2023unsupervised,lyu2022learning,hu2022synthetic,hu2023synthetic} and pancreas~\cite{wei2022pancreatic,li2023early}. AI trained on these synthetic tumors perform similarly well as those trained with real tumors. However, these methods need to be redesigned for tumors in other organs, which severely limits the generalization capabilities. In this work, we learn the tumor distribution based on generative models, i.e., Diffusion Models, to realize generalizable tumor synthesis.

\section{Conclusion}
\label{sec:conclusion}

This work introduces \ourmodel\ for generalizable tumor synthesis. We leverage the observation that early-stage tumors share similar imaging characteristics in CT scans across different organs (e.g., liver, pancreas, kidneys). As a result, \ourmodel\ trained solely on annotated liver tumors can directly synthesize tumors in other organs with limited annotated data (e.g., pancreas, kidney). By augmenting large-scale datasets of healthy organs (readily available in clinical settings) with these synthetic tumors, we substantially expand training data for tumor segmentation models. This augmentation significantly improves AI generalizability across diverse hospital systems and patient populations.

\smallskip\noindent\textbf{Acknowledgments.}
This work was supported by the Lustgarten Foundation for Pancreatic Cancer Research and the Patrick J. McGovern Foundation Award. We thank Yuxiang Lai, Qian Yu, and Wenxuan Li for their constructive suggestions at several stages of the project.

\newpage
{\small
\bibliographystyle{ieee_fullname}
\bibliography{refs,zzhou}
}

\newpage
\clearpage
\appendix

\onecolumn
This appendix is organized as follows: \S\ref{sec:reader_study_examples_appendix} provides visual examples for reader study and Visual Turing Test. \S\ref{sec:radiomics_feature_appendix} provides a description of Radiomics Features. \S\ref{sec:generalizable_organs_appendix} provides additional results for generalizable to multiple organs. \S\ref{sec:generalizable_hospitals_appendix} provides additional results for generalizable to different patient demographics. \S\ref{sec:dataset_implementation_details_appendix} provides the details of used datasets and implementation for \ourmodel\ and Segmentation Model. \S\ref{sec:discussion_appendix} provides discussions about comparison with related works, unrealistic generation, and challenging case analysis.

\section{Visual Examples}
\label{sec:reader_study_examples_appendix}

\begin{figure}[h]
	\centering
	\includegraphics[width=0.95\columnwidth]{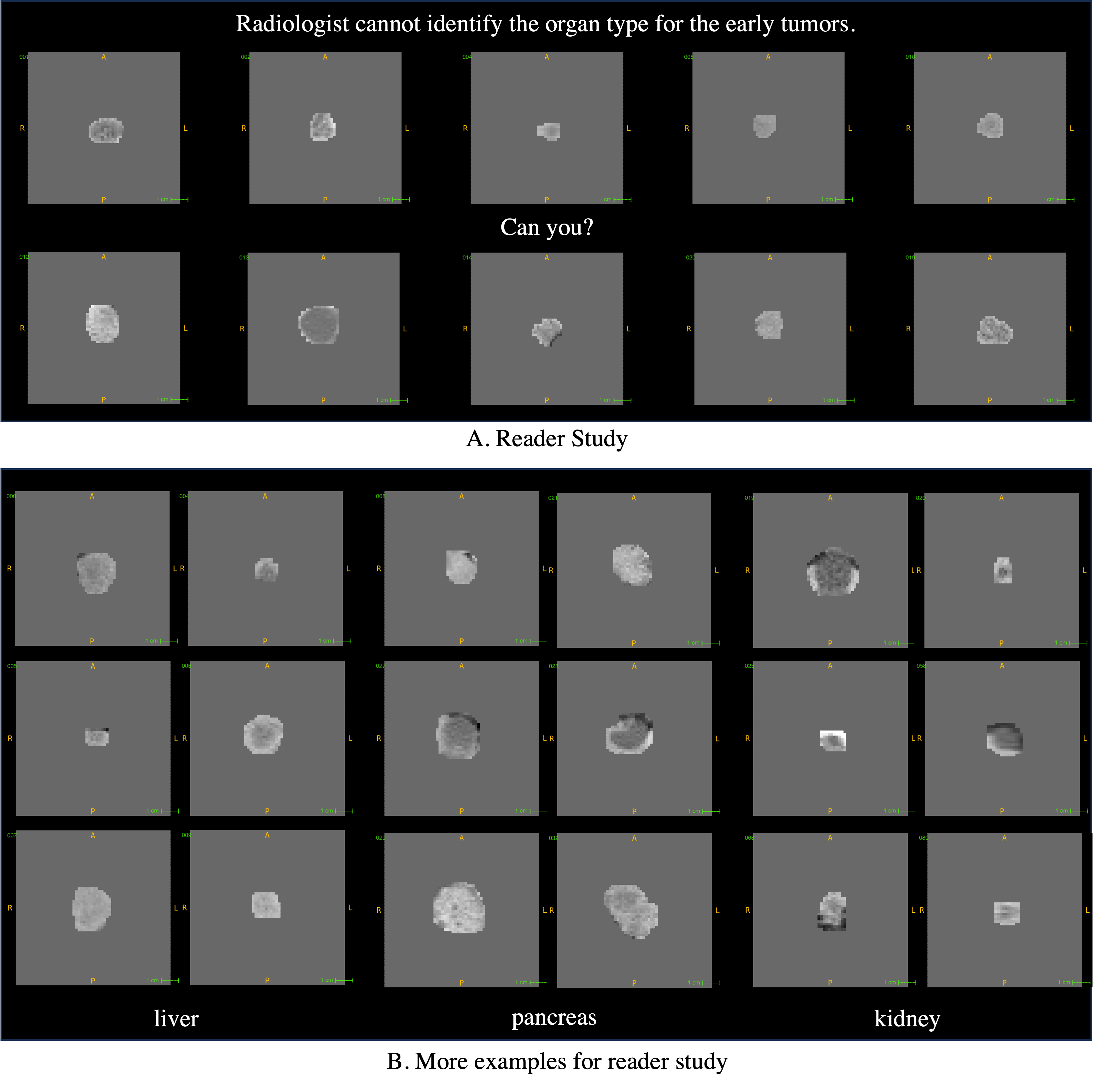}
    \caption{
    \textbf{Visual examples for the reader study.} \textbf{A.} Quick test for identifying the organ type for the early tumors. \textbf{B.} More early tumors for three abdominal organs. The organs corresponding to the early tumor in the first row of \textbf{A} are the liver, pancreas, kidney, liver, and kidney. The organs corresponding to the early tumor in the second row of \textbf{A} are the pancreas, pancreas, liver, liver, and pancreas.
    } 
	\label{fig:supp_reader_study_examples}
\end{figure}

\begin{figure}[h]
	\centering
	\includegraphics[width=0.95\columnwidth]{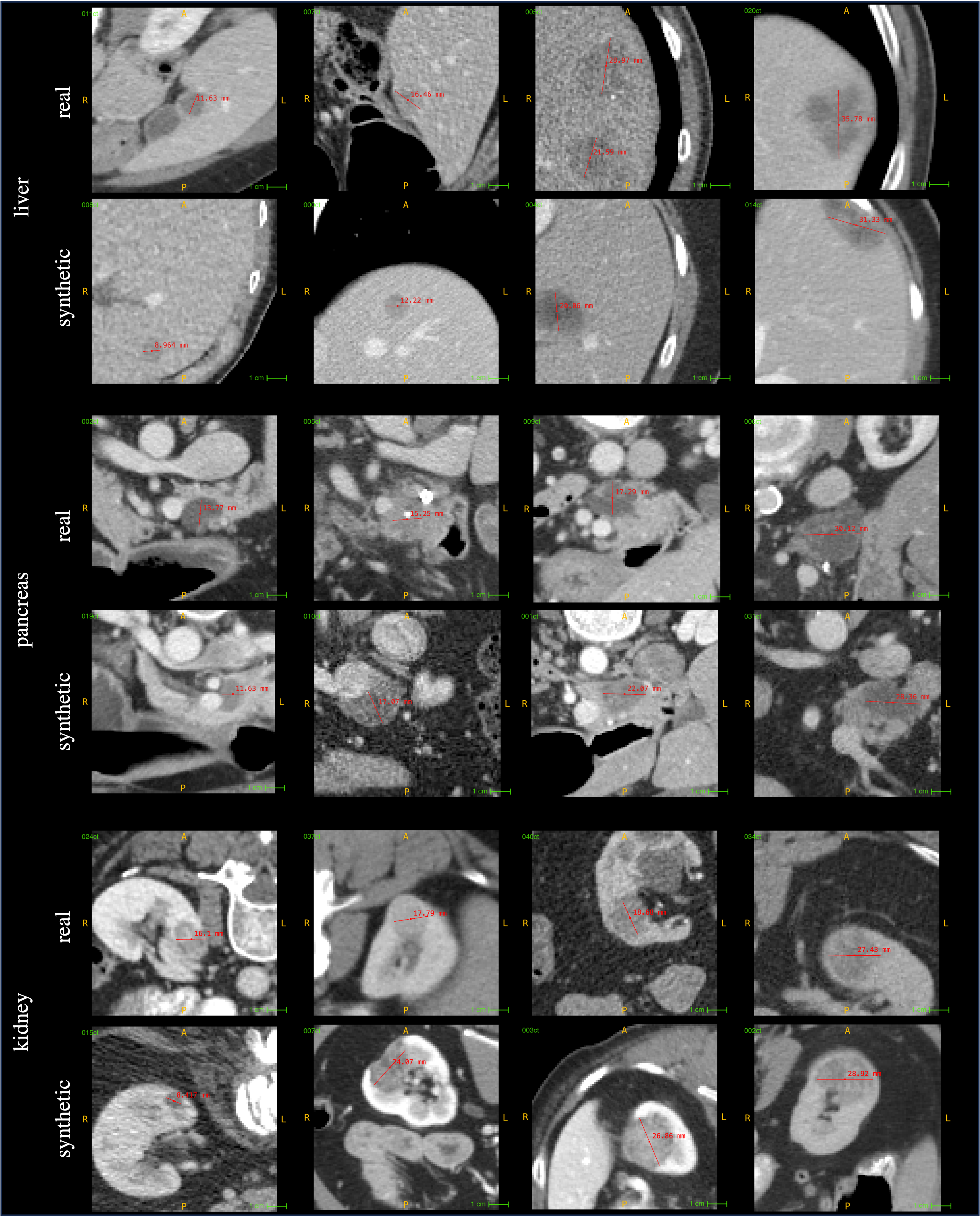}
    \caption{
    \textbf{Visual examples for the Visual Turing Test.} We present real and synthetic tumors, arranged from smaller to larger sizes (columns 1–4), for the Visual Turing Test. The radiologists are instructed to classify each tumor as either real or synthetic. Based on results in \S\ref{sec:visual_turing_test} and Table~\ref{tab:reader_studies},  a minimum of 50\% synthetic tumors are identified as real by both radiologists. 
    } 
	\label{fig:supp_visual_turing_test_examples}
\end{figure}

\clearpage
\section{Description of Radiomics Features}
\label{sec:radiomics_feature_appendix}

Radiomics Features~\cite{van2017computational} consist of a comprehensive set of quantitative, high-dimensional imaging attributes derived from radiographic images, such as Computed Tomography (CT) scans, Magnetic Resonance Imaging (MRI), and Positron Emission Tomography (PET) scans. These attributes capture a broad spectrum of image characteristics, including shape, intensity, texture, and wavelet, among others. The scope of Radiomics Features applications is expansive, ranging from predicting disease prognosis to formulating treatment strategies and evaluating treatment response. The effectiveness of these features has been validated across various medical fields, notably in oncology, neurology, and cardiology.

The key characteristics of Radiomics Features include their high-throughput capacity and reproducibility, which facilitate a detailed characterization of tumor phenotypes. These features are multivariate, incorporating first-order statistics, shape and size-based features, textural features, and filter-based features.

In this paper, we utilize the official Radiomics feature repository\footnote{\href{https://github.com/AIM-Harvard/pyradiomics/}{https://github.com/AIM-Harvard/pyradiomics/}} to extract the appearance features, which include 3D shape-based features (16 dimensions), gray level co-occurrence matrix (24 dimensions), gray level run length matrix (16 dimensions), gray level size zone matrix (16 dimensions), neighboring gray-tone difference matrix (5 dimensions), and gray level dependence matrix (14 dimensions). The shape descriptors are independent of the gray value and are extracted from the tumor mask. The definitions of these features can be referenced at \href{https://pyradiomics.readthedocs.io/en/latest/features.html}{https://pyradiomics.readthedocs.io/en/latest/features.html}. Based on the tumor mask annotations, we are able to extract only the appearance features of tumors. Consequently, for each early-stage tumor, a 91-dimensional vector can be obtained. Ultimately, these features from all early-stage tumors are aggregated for Radiomics feature analysis in \figureautorefname~\ref{fig:preliminary}.

\clearpage
\section{Generalizable to Multiple Organs}
\label{sec:generalizable_organs_appendix}

\begin{table}[h]
    \centering
    \footnotesize
    \begin{tabular}{p{0.14\linewidth}p{0.16\linewidth}P{0.15\linewidth}P{0.15\linewidth}P{0.15\linewidth}}
    \multicolumn{5}{l}{\textbf{nnU-Net}~\cite{isensee2021nnu}} \\
    \toprule
    \multicolumn{2}{l}{source $\backslash$ target} & liver & pancreas & kidneys \\
    \midrule
    \multirow{3}{*}{\makecell[l]{liver}} & real tumors & 77.4 &1.8 & 2.4\\
     & Hu~\etal~\cite{hu2023label} &78.0  &56.3  &61.9 \\
     & \ourmodel &\cellcolor{iblue!10}\textbf{80.9}  &\cellcolor{iblue!10}\textbf{60.7} &\cellcolor{iblue!10}\textbf{71.4} \\
    \midrule
    \multirow{3}{*}{\makecell[l]{pancreas}} & real tumors &1.5  &  67.0 &2.4 \\
     & Hu~\etal~\cite{hu2023label} &\cellcolor{iblue!10}\textbf{76.2}  &68.8  &61.9 \\
     & \ourmodel &72.8 &\cellcolor{iblue!10}\textbf{75.0}  &\cellcolor{iblue!10}\textbf{81.0} \\
    \midrule
    \multirow{3}{*}{\makecell[l]{kidney}} & real tumors &0.9   &0.9   &  59.5 \\
     & Hu~\etal~\cite{hu2023label} &76.2  &56.3  &69.0 \\
     & \ourmodel &\cellcolor{iblue!10}\textbf{77.4}  &\cellcolor{iblue!10}\textbf{63.4}  &\cellcolor{iblue!10}\textbf{76.2} \\
    \bottomrule
    \end{tabular} 
    \begin{tabular}{p{0.14\linewidth}p{0.16\linewidth}P{0.15\linewidth}P{0.15\linewidth}P{0.15\linewidth}} \\
    \multicolumn{5}{l}{\textbf{Swin UNETR}~\cite{hatamizadeh2021swin}} \\
    \toprule
    \multicolumn{2}{l}{source $\backslash$ target} & liver & pancreas & kidneys \\
    \midrule
    \multirow{3}{*}{\makecell[l]{liver}} & real tumors & 76.2 &2.7  &0  \\
     & Hu~\etal~\cite{hu2023label} &79.2  &63.4  &71.4 \\
     & \ourmodel &\cellcolor{iblue!10}\textbf{83.1}  &\cellcolor{iblue!10}\textbf{69.6} &\cellcolor{iblue!10}\textbf{73.8} \\
    \midrule
    \multirow{3}{*}{\makecell[l]{pancreas}} & real tumors &1.6   &  70.5 &4.8  \\
     & Hu~\etal~\cite{hu2023label} &66.4  &73.2  &71.4 \\
     & \ourmodel &\cellcolor{iblue!10}\textbf{76.2} &\cellcolor{iblue!10}\textbf{79.5}  &\cellcolor{iblue!10}\textbf{90.5} \\
    \midrule
    \multirow{3}{*}{\makecell[l]{kidney}} & real tumors &0.6   &0  &  69.0 \\
     & Hu~\etal~\cite{hu2023label} &66.4  &\cellcolor{iblue!10}\textbf{63.4}  &76.2 \\
     & \ourmodel &\cellcolor{iblue!10}\textbf{76.2}  &61.6  &\cellcolor{iblue!10}\textbf{81.0}  \\
    \bottomrule
    \end{tabular}
    \caption{
    \textbf{Generalizable across various organs.} We show the comparison of generalization for early-stage tumor detection (measured in tumor-wise Sensitivity \%) using additional backbones. The scores highlighted in bold denote the superior performance in each respective domain. Consistent with the results on U-Net presented in \tableautorefname~\ref{tab:generalizable_organs}, \ourmodel\ demonstrates superior performance across nearly all domains on nnU-Net and Swin UNETR.
    }
    \label{tab:supp_generalizable_organs}
\end{table}

\begin{table*}[!ht]
    \centering
    \footnotesize
    \begin{tabular}{p{0.15\linewidth}p{0.09\linewidth}P{0.08\linewidth}P{0.08\linewidth}P{0.08\linewidth}P{0.08\linewidth}P{0.08\linewidth}P{0.16\linewidth}P{0.08\linewidth}}
        \toprule
        \textbf{U-Net} &metrics & fold0 & fold1 & fold2 & fold3 & fold4 & average\\
        \midrule
        \multirow{2}{*}{real tumors}  & DSC (\%)  &62.3  &72.8  &63.8 &54.5 &59.0 & 62.5\\
          &NSD (\%)  &63.4  &74.6 & 63.3 & 55.5 &61.6 & 63.7\\ \midrule
         \multirow{2}{*}{\ourmodel} & DSC (\%)  &70.9 &74.0  &67.9&59.1&60.6 &66.5 \\
         &NSD (\%)  &71.2  &73.9&70.4 &61.0&63.3&68.0\\  \midrule \\
        \textbf{nnU-Net} &metrics & fold0 & fold1 & fold2 & fold3 & fold4 & average\\
        \midrule
        \multirow{2}{*}{real tumors}  & DSC (\%)  &64.3 & 70.2 & 64.3 & 56.3 & 59.6 & 62.9\\
       &NSD (\%) &65.7 & 72.7& 63.1& 59.3& 62.6 & 64.7\\ \midrule
         \multirow{2}{*}{\ourmodel}  & DSC (\%)  &73.6  &73.9  &67.6&64.9&63.8 &68.8 \\
          &NSD (\%)  &75.3  &73.9&67.9&69.0&66.5&70.5 \\
        \midrule \\
        \textbf{Swin UNETR} &metrics & fold0 & fold1 & fold2 & fold3 & fold4 & average\\
        \midrule
        \multirow{2}{*}{real tumors}  & DSC (\%)  &65.1  &69.4 &57.4 &59.0 &58.2 & 61.8\\
          &NSD (\%) &65.9 & 71.7 &53.1 & 62.1 &61.9 &62.8\\ \midrule
         \multirow{2}{*}{\ourmodel}  & DSC (\%)  &71.4  &71.7  &71.6 &62.2 &62.4&67.9 \\
          &NSD (\%)  &73.5  &72.4 &74.5 &66.5 &66.0&70.6  \\
        \bottomrule
        \multicolumn{8}{l}{\textit{real tumors denotes Segmentation Model trained on 95 CT scans containing real tumors.}} \\
    \multicolumn{8}{l}{\textit{\ourmodel\ denotes Segmentation Model trained on 95 CT scans containing real tumors and 116 healthy CT scans.}}
    \end{tabular}
        \caption{\textbf{Liver tumor segmentation performance on 5-fold cross-validation.} We conduct a comparative analysis of the Segmentation Model (U-Net, nnU-Net, Swin UNETR) trained on both synthetic and real tumors against the model trained exclusively on real tumors, employing 5-fold cross-validation. The evaluation metrics employed include the Dice Similarity Coefficient (DSC) and the Normalized Surface Distance (NSD). \ourmodel\ consistently enhances liver tumor segmentation performance across these three backbones.
    }
    \label{tab:liver_5fold_result}
\end{table*}

\begin{table*}[!ht]
    \centering
    \footnotesize
    \begin{tabular}{p{0.15\linewidth}p{0.09\linewidth}P{0.08\linewidth}P{0.08\linewidth}P{0.08\linewidth}P{0.08\linewidth}P{0.08\linewidth}P{0.16\linewidth}P{0.08\linewidth}}
        \toprule
        \textbf{U-Net} &metrics & fold0 & fold1 & fold2 & fold3 & fold4 & average\\
        \midrule
        \multirow{2}{*}{real tumors}   & DSC (\%)  &56.0 &51.9&45.5&59.4&43.2&51.2\\
       &NSD (\%)  &51.0&49.9&43.6&57.7&40.2&48.5\\ \midrule
         \multirow{2}{*}{\ourmodel}  & DSC (\%)  &64.8  &58.0  &57.7 &67.9 &51.8 & 60.0\\
          &NSD (\%)  &60.5  &55.3&58.3 &67.5&47.9 &57.9\\
        \midrule \\
        \textbf{nnU-Net} &metrics & fold0 & fold1 & fold2 & fold3 & fold4 & average\\
        \midrule
        \multirow{2}{*}{real tumors}   & DSC (\%)  &59.9 &50.0&44.8&63.5&50.4 &53.7\\
          &NSD (\%) &55.7&47.0&47.0&62.3&48.0&52.0\\ \midrule
         \multirow{2}{*}{\ourmodel}  & DSC (\%)  &63.6  &60.5  &62.5&67.8&55.3&61.9  \\
          &NSD (\%)  &61.1  &59.1&63.4&67.7&54.9&61.2 \\
        \midrule \\
        \textbf{Swin UNETR} &metrics & fold0 & fold1 & fold2 & fold3 & fold4 & average\\
        \midrule
        \multirow{2}{*}{real tumors} & DSC (\%)  &52.2&49.5&50.9&60.0&52.1&52.9\\
          &NSD (\%) &49.1&49.7&50.9&58.9&47.4&51.2\\  \midrule
         \multirow{2}{*}{\ourmodel}  & DSC (\%)  &62.2  &60.2  &59.0 &69.7 &53.8&61.0  \\
          &NSD (\%)  &58.7  &58.4 &62.8 &67.2 &51.2&59.7\\
        \bottomrule
    \multicolumn{8}{l}{\textit{real tumors denotes Segmentation Model trained on 96 CT scans containing real tumors.}} \\
    \multicolumn{8}{l}{\textit{\ourmodel\ denotes Segmentation Model trained on 96 CT scans containing real tumors and 120 healthy CT scans.}}
    \end{tabular}
    
        \caption{\textbf{Pancreatic tumor segmentation performance on 5-fold cross-validation.} We execute a comparative study of the Segmentation Model (U-Net, nnU-Net, Swin UNETR) trained on both synthetic and real tumors against the model trained exclusively on real tumors, utilizing 5-fold cross-validation. The employed evaluation metrics are the Dice Similarity Coefficient (DSC) and the Normalized Surface Distance (NSD). \ourmodel consistently yields a significant improvement in pancreatic tumor segmentation across these three backbones. It should be noted that the segmentation of pancreatic tumors is deemed the most challenging task among the three abdominal organs in study. The enhancement observed in pancreatic tumor segmentation is the most substantial among the three.
    }
    \label{tab:pancreatic_5fold_result}
\end{table*}

\begin{table*}[!ht]
    \centering
    \footnotesize
    \begin{tabular}{p{0.15\linewidth}p{0.09\linewidth}P{0.08\linewidth}P{0.08\linewidth}P{0.08\linewidth}P{0.08\linewidth}P{0.08\linewidth}P{0.16\linewidth}P{0.08\linewidth}}
        \toprule
        \textbf{U-Net} &metrics & fold0 & fold1 & fold2 & fold3 & fold4 & average\\
        \midrule
        \multirow{2}{*}{real tumors}  & DSC (\%)  &75.1&68.0&69.0&78.1&70.6&72.0\\
         &NSD (\%)  &68.4&59.0&57.7&68.3&62.4&63.2\\  \midrule
         \multirow{2}{*}{\ourmodel}  & DSC (\%)  &84.2  &76.7  &79.4  &80.6 &74.1 & 79.0\\
          &NSD (\%)  &76.6  &64.5 &70.7&71.7&65.8 &69.9\\
        \midrule \\
        \textbf{nnU-Net} &metrics & fold0 & fold1 & fold2 & fold3 & fold4 & average\\
        \midrule
        \multirow{2}{*}{real tumors} & DSC (\%)  &73.8&76.8&80.0&80.5&73.4&76.9\\
          &NSD (\%) &62.7&70.2&71.2&70.8&67.5&68.5\\ \midrule
         \multirow{2}{*}{\ourmodel} & DSC (\%)  &84.5  &83.4  &81.6 &83.9  &77.3&82.1\\
          &NSD (\%)  &78.3  &74.4 &74.1&76.9&72.3&75.2\\
        \midrule \\
        \textbf{Swin UNETR} &metrics & fold0 & fold1 & fold2 & fold3 & fold4 & average\\
        \midrule
        \multirow{2}{*}{real tumors}   & DSC (\%)  &80.6&64.9&79.1&76.0&72.2&74.6\\
          &NSD (\%) &74.2&55.5&68.2&67.4&64.8&66.0\\ \midrule
         \multirow{2}{*}{\ourmodel} & DSC (\%)  &85.1  &77.2  &81.2&85.7 &79.9&81.8  \\
          &NSD (\%)  &79.2  &70.8&74.1&78.0&74.1&75.2 \\
        \bottomrule
        \multicolumn{8}{l}{\textit{real tumors denotes Segmentation Model trained on 96 CT scans containing real tumors.}} \\
    \multicolumn{8}{l}{\textit{\ourmodel\ denotes Segmentation Model trained on 96 CT scans containing real tumors and 120 healthy CT scans.}}
    \end{tabular}
        \caption{\textbf{Kidney tumor segmentation performance on 5-fold cross-validation.} We perform a comparative analysis of the Segmentation Model (U-Net, nnU-Net, Swin UNETR) trained on both synthetic and real tumors versus the model trained exclusively on real tumors, employing 5-fold cross-validation. Our evaluation metrics include the Dice Similarity Coefficient (DSC) and the Normalized Surface Distance (NSD). Similar to the other two tumor segmentation tasks, \ourmodel\ can deliver substantial improvements in kidney tumor segmentation across these three prevalent backbones.
    }
    \label{tab:kidney_5fold_result}
\end{table*}

\begin{figure}[t]
	\centering
	\includegraphics[width=\columnwidth]{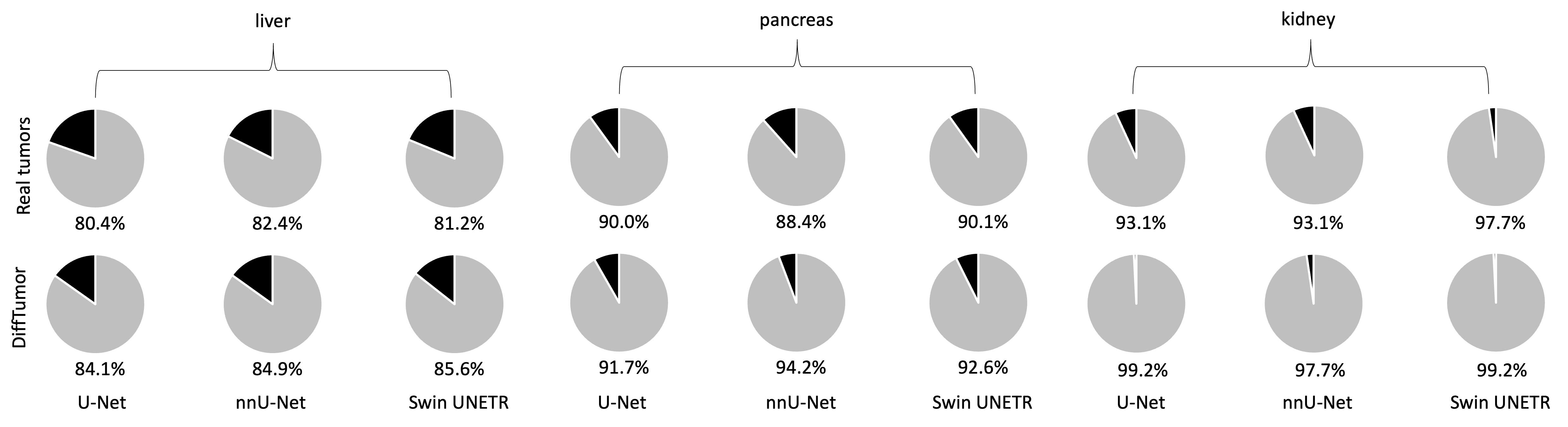}
	\caption{
    \textbf{Enhancement in all-stage tumor detection.} We present the tumor detection results, measured by tumor-wise sensitivity, across different Segmentation Models (U-Net, nnU-Net, Swin UNETR). Consistent with the results in tumor segmentation performance, \ourmodel\ can significantly enhance tumor detection performance across these three common backbones.
    } 
	\label{fig:supp_augmentation_sensitivity}
\end{figure}

\begin{figure}[t]
	\centering
	\includegraphics[width=\columnwidth]{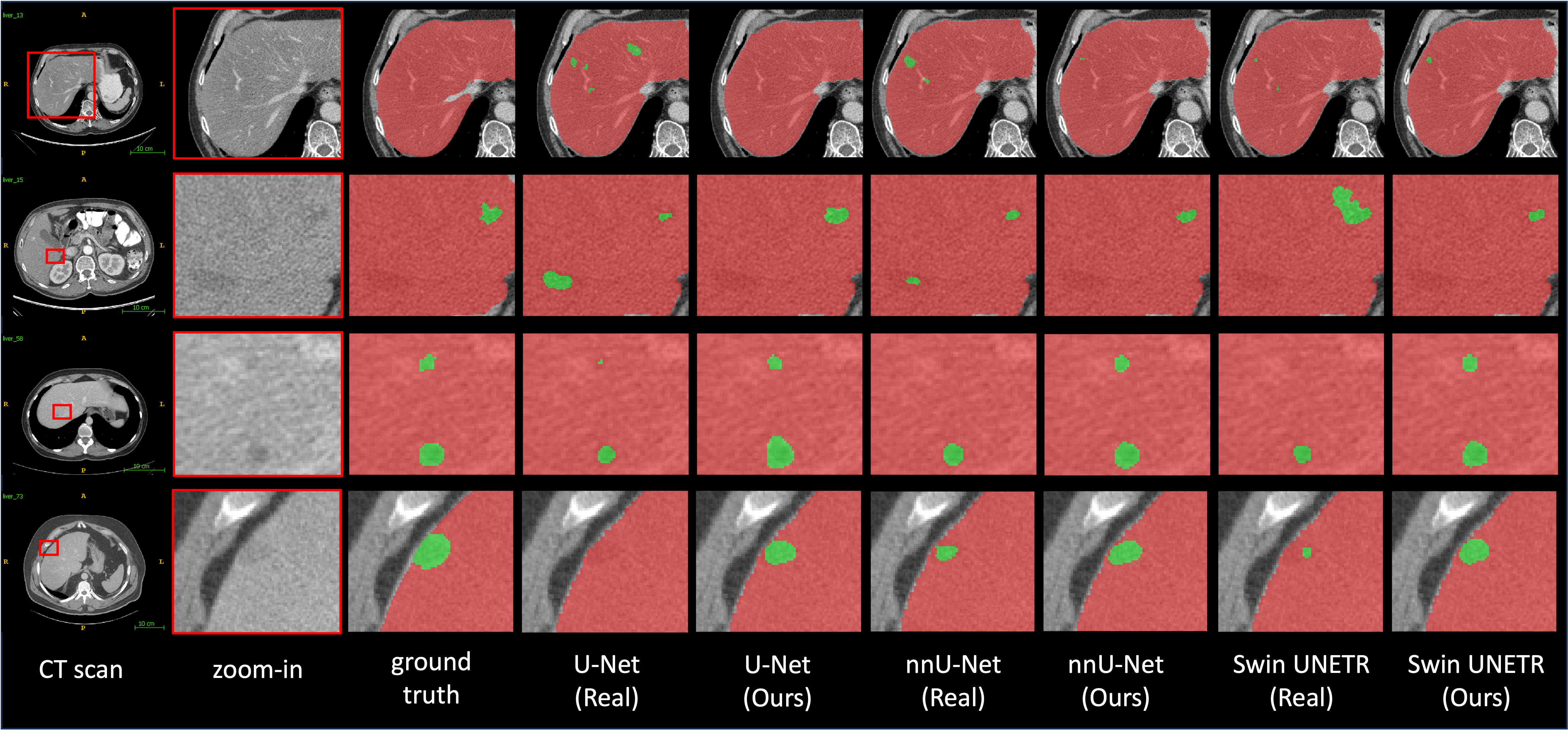}
 \caption{
    \textbf{Liver tumor segmentation.} We provide qualitative visualizations of Segmentation Models (U-Net, nnU-Net, Swin UNETR) for liver tumor segmentation. The results from the first and second rows illustrate that \ourmodel\ can effectively reduce the number of false positive cases to a certain extent. The results in the third and fourth rows indicate that \ourmodel\ can improve the detection rate within the tumor region. Consequently, our method yields a substantial improvement in segmentation performance, as evidenced in \tableautorefname~\ref{tab:liver_5fold_result}.
    } 
	\label{fig:supp_liver_tumor_detection}
\end{figure}

\begin{figure}[t]
	\centering
	\includegraphics[width=\columnwidth]{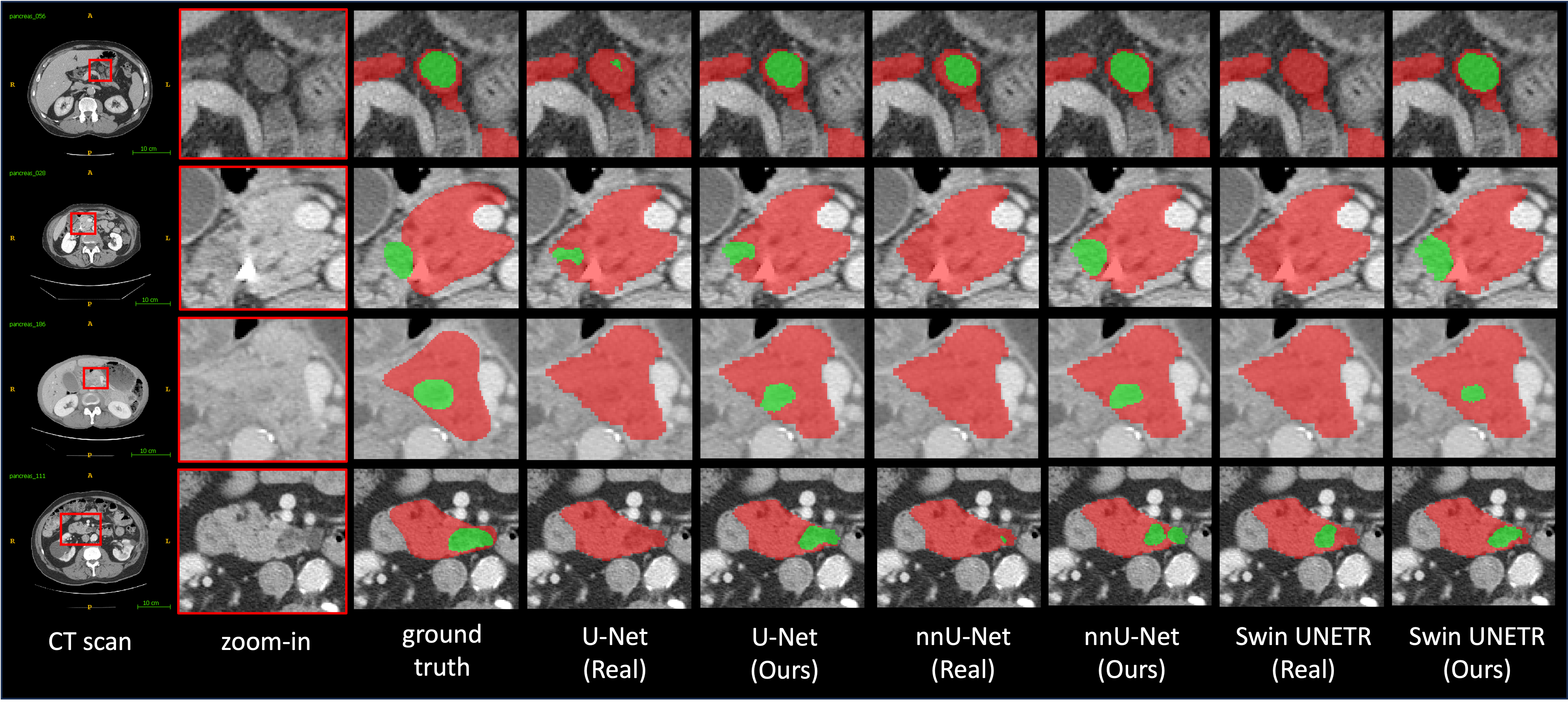}
    \caption{
    \textbf{Pancreatic tumor segmentation.} We provide qualitative visualizations of Segmentation Models (U-Net, nnU-Net, Swin UNETR) for pancreatic tumor segmentation. Pancreatic tumor segmentation is more challenging as many tumors are easily missed by Segmentation Models. The results displayed in rows 1-4 indicate that \ourmodel\ can aid detecting tumors that were missed during training on real tumors, thereby resulting in a significant improvement in segmentation performance, as illustrated in \tableautorefname~\ref{tab:pancreatic_5fold_result}.	
    } 
	\label{fig:supp_pancreatic_tumor_detection}
\end{figure}

\begin{figure}[t]
	\centering
	\includegraphics[width=\columnwidth]{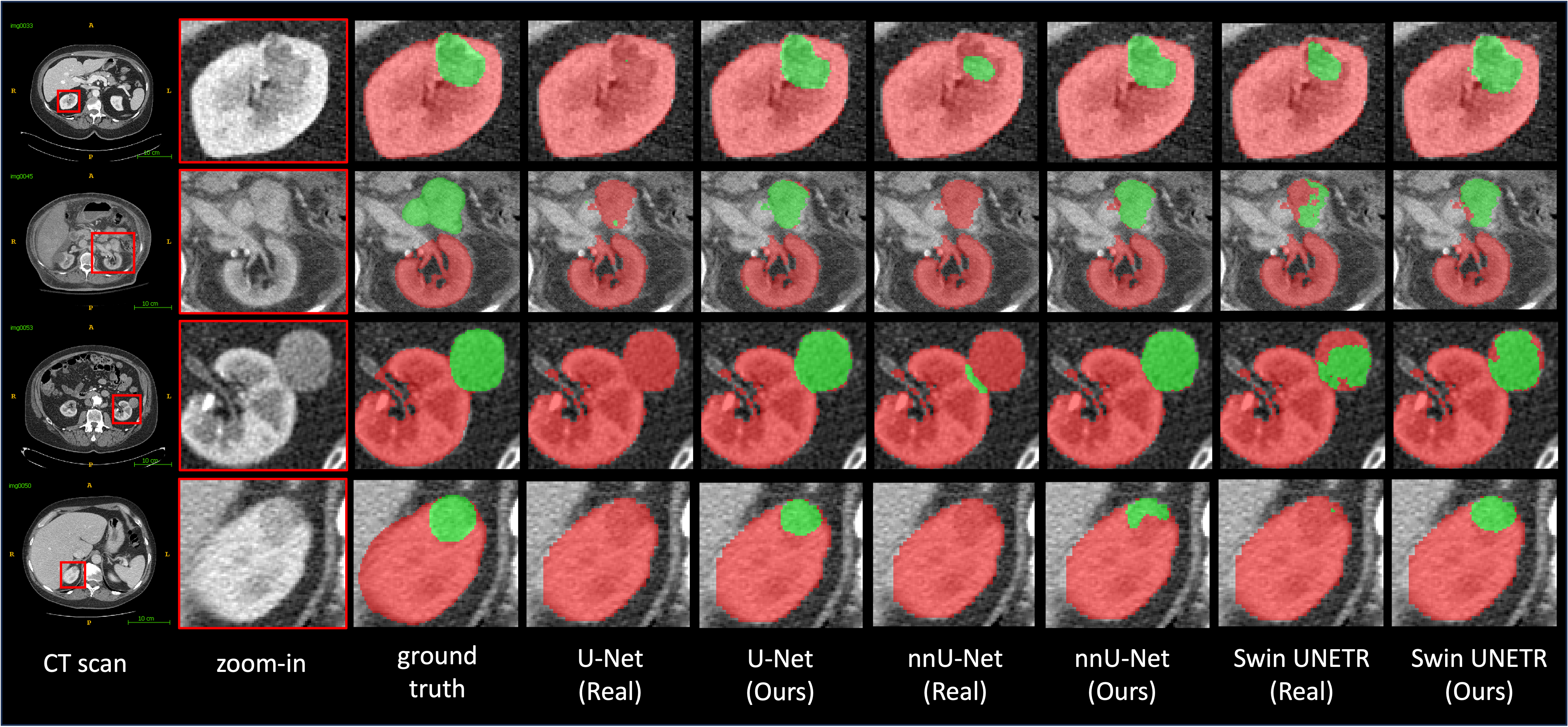}
    \caption{
    \textbf{Kidney tumor segmentation.} We provide qualitative visualizations of Segmentation Models (U-Net, nnU-Net, Swin UNETR) for kidney tumor segmentation. Segmentation Models, when trained on real tumors, will miss the tumors located on the boundary of the kidney or outside the kidney. \ourmodel\ can help enhance the segmentation of these difficult cases, thereby improving the overall segmentation performance, as evidenced in \tableautorefname~\ref{tab:kidney_5fold_result}.
    } 
	\label{fig:supp_kidney_tumor_detection}
\end{figure}

\clearpage
\section{Generalizable to Different Patient Demographics}
\label{sec:generalizable_hospitals_appendix}

\begin{figure}[h]
	\centering
	\includegraphics[width=\columnwidth]{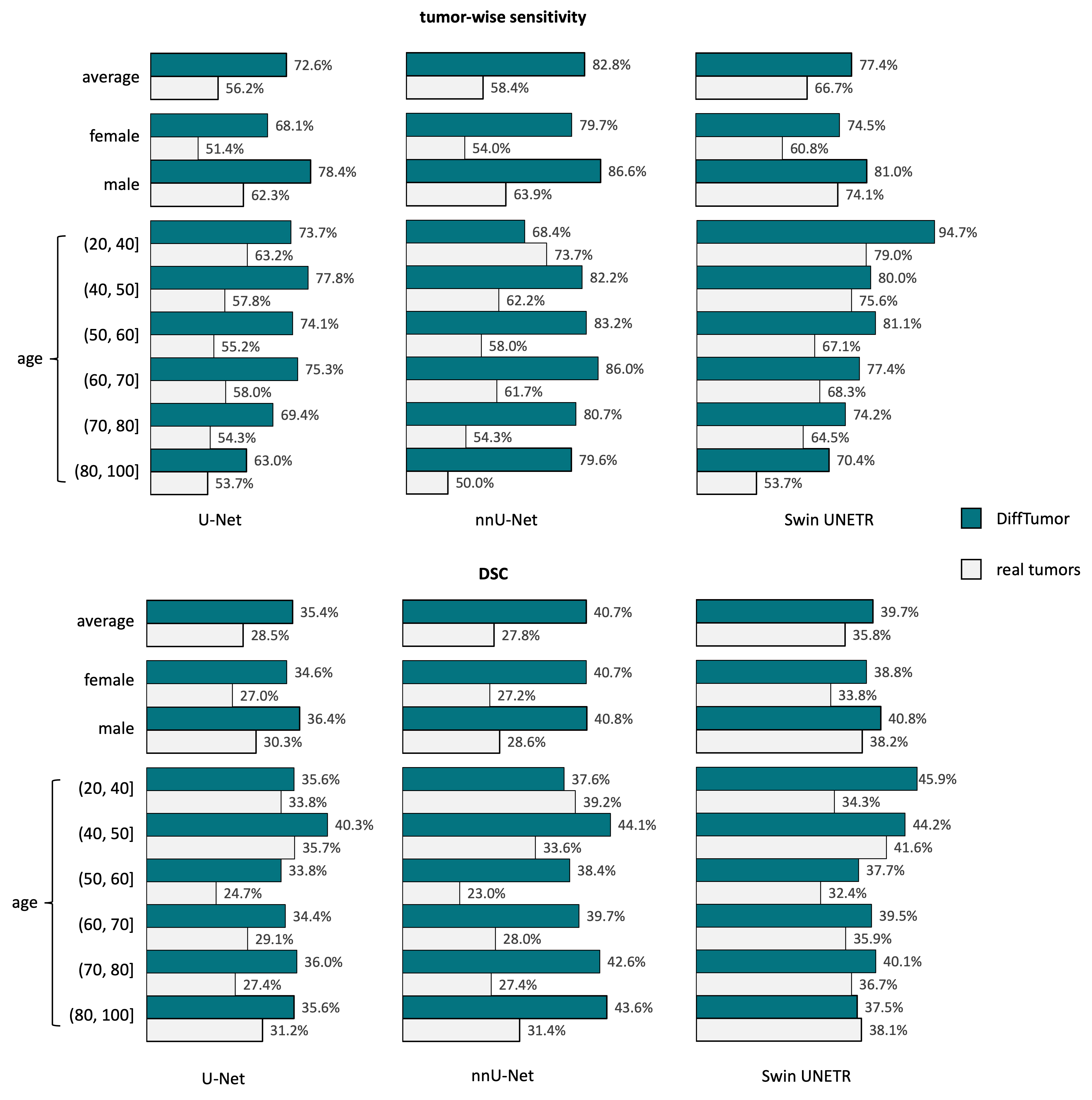}
    \caption{
    \textbf{Generalizable across different patient demographics.} A comparison of generalization across different patient demographics for tumor detection (tumor-wise Sensitivity \%) and segmentation (DSC \%) using different backbones is provided. \ourmodel\ can consistently enhance tumor detection and segmentation performance by a significant margin across various patient groups and different backbones. The most significant improvement is observed on nnU-Net.
    } 
	\label{fig:supp_generalizable_hospitals}
\end{figure}

\clearpage
\section{Dataset \& Implementation Details}
\label{sec:dataset_implementation_details_appendix}
\subsection{Dataset Details}
\smallskip\noindent\textbf{\textit{Real-tumor datasets.}} LiTS~\citep{bilic2019liver} comprises 131 and 70 contrast-enhanced 3-D abdominal CT scans for training and testing, respectively. This dataset was compiled utilizing various scanners and protocols from six unique clinical sites, which resulted in a significant variation in in-plane resolution (ranging from 0.55 to 1.0 mm) and slice spacing (ranging from 0.45 to 6.0 mm). The MSD-Pancreas~\cite{antonelli2021medical} dataset comprises 420 portal-venous phase CT scans from patients who underwent pancreatic mass resection. It includes 281 CT scans designated for training and 139 CT scans for testing. The annotations provided correspond to the pancreatic parenchyma and pancreatic mass. KiTS~\cite{heller2020international} includes 210 CT scans for training and 90 CT scans for testing. Each CT scan features one or more kidney tumors. The University of Minnesota Medical Center provides the annotations.

\smallskip\noindent\textbf{\textit{Healthy-organ datasets.}}
AbdomenAtlas-8K~\cite{qu2023annotating} is currently the most extensive multi-organ dataset, with annotations for the spleen, liver, kidneys, stomach, gallbladder, pancreas, aorta, and IVC in 8,448 CT volumes, which equates to 3.2 million slices. AbdomenAtlas-8K consolidates datasets from 26 distinct hospitals worldwide. In this study, we utilize the CLIP-Driven Universal Model~\cite{liu2023clip,zhang2023continual} to select CT scans that feature the corresponding healthy abdominal organs. This model, ranking first in the Medical Segmentation Decathlon (MSD) competition, has demonstrated high sensitivity and specificity in tumor detection. Consequently, we employ the pre-trained weights\footnote{\href{https://github.com/ljwztc/CLIP-Driven-Universal-Model}{https://github.com/ljwztc/CLIP-Driven-Universal-Model}} using the Swin UNETR backbone to identify CT scans where the prediction includes the organs but does not include the corresponding tumors. Through this process, we have obtained 1246 CT volumes with healthy livers, 1901 CT volumes with healthy pancreas, and 1005 CT volumes with healthy kidneys.

\smallskip\noindent\textbf{\textit{Proprietary dataset.}}
It comprises 5,038 CT scans with 21 annotated organs, with each case having been scanned by contrast-enhanced CT in both venous and arterial phases, utilizing Siemens MDCT scanners. In this study, we utilize 532 CT scans containing 690 PDAC to assess the generalizability of the Segmentation Model across varied patient demographics. The test set we used includes 243 CT scans of males and 289 CT scans of females. Additionally, the age range of these patients spans from 20 to 146, essentially covering all stages of a person's life.

\subsection{Implementation Details}
\smallskip\noindent\textbf{\textit{Autoencoder Model.}}
In this study, we train Autoencoder Model on a total of 9262 CT scans from the AbdomenAtlas-8K dataset and a private dataset. The purpose is to learn a general low-dimensional latent representation of CT scans. The model processes the CT volume into a latent feature, reducing the original input volume's size by 1/4 in height, width, and depth, respectively. We set the codebook size and dimensionality at 16384 and 8, respectively. CT scan orientation is adjusted to specific axcodes, and isotropic spacing is applied to resample each scan, resulting in a uniform voxel size of $1.0 \times 1.0 \times 1.0 mm^3$. Additionally, the intensity in each scan is truncated to the range $\left[−175, 250\right]$ and then linearly normalized to $\left[-1, 1\right]$. During training, we crop random fixed-sized $96\times96\times96$ regions. We employ the Adam optimizer for training with $\beta_1$ and $\beta_2$ hyperparameters set to 0.9 and 0.999, respectively, a learning rate of 0.0003, and a batch size of 4 per GPU. The training is conducted over a week on a node with four A100 GPUs, completing 200k iterations.

\smallskip\noindent\textbf{\textit{Diffusion Model.}} In this study, we train the corresponding Diffusion Model specifically for tumors of three different abdominal organs. The data preprocessing carried out during the training phase is identical to the approach used for training Autoencoder Model.  Besides, we utilize the Adam optimizer for training with $\beta_1$ and $\beta_2$ hyperparameters set to 0.9 and 0.999, respectively, a learning rate of 0.0001, and a batch size of 10 per GPU.  The training is conducted over the course of a day on a node with an A100 GPU for 60k iterations.

\smallskip\noindent\textbf{\textit{Segmentation Model.}}
The code for the Segmentation Model is implemented in Python using MONAI\footnote{\href{https://monai.io/}{https://monai.io/}}. In this study, we implement Swin UNETR based on the Swin UNETR Base variant. The orientation of CT scans is adjusted to specific axcodes. Isotropic spacing is utilized to resample each scan to achieve a uniform voxel size of $1.0 \times 1.0 \times 1.0 mm^3$. Besides, the intensity in each scan is truncated to the range $\left[−175, 250\right]$ and then linearly normalized to $\left[0, 1\right]$. During training, we crop random ﬁxed-sized $96\times96\times96$ regions with the center being a foreground or background voxel based on the predeﬁned ratio. Additionally, the input patch is randomly rotated by 90 degrees, and the intensity is shifted with a 0.1 offset, each with probabilities of 0.1 and 0.2, respectively. To avoid confusion between the organs on the right and left sides, mirroring augmentation is not employed. All models on real tumors are trained for 3,000 epochs and models on synthetic and real tumors are trained for 2,000 epochs. Moreover, the base learning rate is set at 0.0002, and the batch size is set at two. We adopt the linear warmup strategy and the cosine annealing learning rate schedule. For details on the tumor synthesis process during the training of the Segmentation Model, please refer to the provided code. For inference, we use the sliding window strategy by setting the overlapping area ratio to 0.75. Besides, to rule out tumor mask predictions that do not belong to the respective organs, we use the pseudo labels of organs obtained through~\cite{liu2023clip} to process the predictions of Segmentation Models.

\section{Discussion}
\label{sec:discussion_appendix}
\subsection{Comparison with Related Works}
In recent works, Hu~\etal~\cite{hu2023label} have synthesized tumors in the liver using a model-based approach. This approach, guided by radiologists, involves several image-processing operations such as ellipse generation, elastic deformation, salt-noise generation, Gaussian filtering, scaling, and clipping. The synthetic tumors are realistic in comparison to real liver tumors. Notably, the AI trained with synthetic tumors achieves segmentation/detection performance that is comparable to the performance of the AI trained with real tumors. 

However, the approach of Hu~\etal~\cite{hu2023label} requires significant effort and expertise to identify the proper imaging characteristics of tumors. In other words, the resulting synthetic tumors need to be explicitly specified by radiologists, tailored to the specific types of tumors and must be redesigned for tumors in other organs. To demonstrate the superiority of \ourmodel\ for enhancing tumor segmentation in the three abdominal organs, we compare it with the representative tumor synthetic strategy by Hu~\etal~\cite{hu2023label}. The results can be found in \tableautorefname~\ref{tab:supp_generalizable_organs}.

\begin{table}[h]
    \centering
    \footnotesize
    \begin{tabular}{p{0.16\linewidth}P{0.11\linewidth}P{0.11\linewidth}P{0.11\linewidth}P{0.11\linewidth}P{0.11\linewidth}P{0.11\linewidth}} 
    \toprule
     \multirow{2}{*}{Methods}& \multicolumn{2}{c}{liver} & \multicolumn{2}{c}{pancreas} & \multicolumn{2}{c}{kidneys} \\ \cmidrule{2-7}
     & DSC (\%) & NSD (\%) & DSC (\%) & NSD (\%) & DSC (\%) & NSD (\%)\\
    \midrule
     real tumors &62.3  &  63.4 &56.0 &51.0&75.1&68.4 \\
      Hu~\etal~\cite{hu2023label} &69.7 &70.9 &55.9 &49.9 &80.8 &71.0\\
     \ourmodel &\cellcolor{iblue!10}\textbf{70.9}&\cellcolor{iblue!10}\textbf{71.2} &\cellcolor{iblue!10}\textbf{64.8}&\cellcolor{iblue!10}\textbf{60.5}&\cellcolor{iblue!10}\textbf{84.2}&\cellcolor{iblue!10}\textbf{76.6}\\
    \bottomrule
    \end{tabular} 
    \caption{
    \textbf{Comparison for tumor segmentation enhancement.} The comparison for all-stage tumor segmentation is conducted based on the U-Net backbone. While Hu~\etal~\cite{hu2023label} is designed for liver tumor synthesis, \ourmodel\ can bring more significant improvement in liver tumor segmentation. The synthesized tumors by Hu~\etal~\cite{hu2023label} can also boost the DSC and NSD scores for kidney tumor segmentation. However, \ourmodel\ can yield better results. Additionally, when adding the synthesized tumors by Hu\etal\cite{hu2023label} for pancreatic tumor segmentation, the DSC and NSD scores even drop compared with training solely on real tumors. This suggests that the synthetic strategy may not be suitable for the synthesis of pancreatic tumors. On the contrary, \ourmodel\ can significantly improve pancreatic tumor segmentation. These results underline the superiority of \ourmodel\ in enhancing tumor segmentation across various abdominal organs.
    }
    \label{tab:supp_generalizable_organs}
\end{table}

\subsection{Unrealistic Generation}
\begin{figure}[h]
	\centering
	\includegraphics[width=\columnwidth]{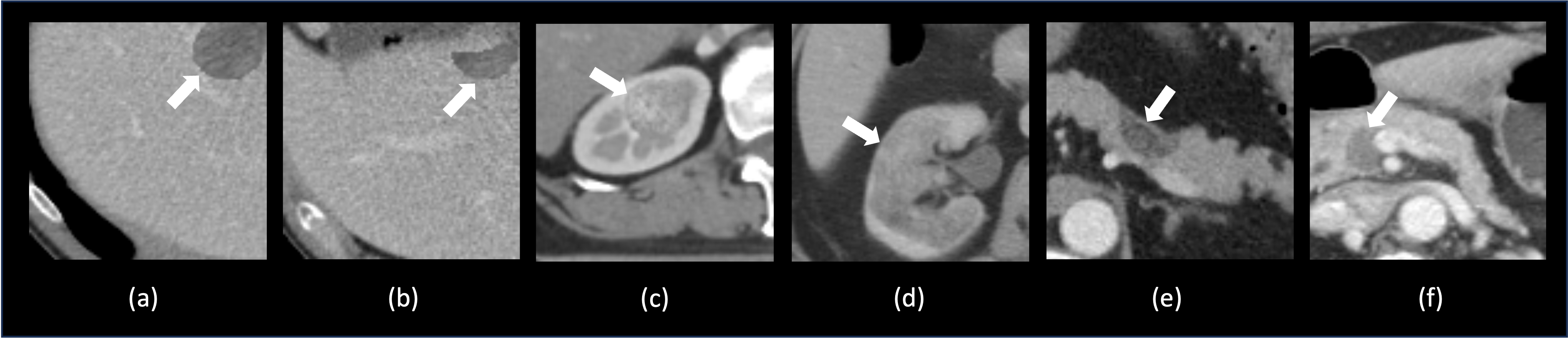}
    \caption{
    \textbf{Unrealistic generation cases.} \textbf{(a)} A synthetic liver tumor with an edge that is too sharp for a malignant tumor, and the inner noise has a discrepancy from the surrounding tissue. \textbf{(b)} A synthetic tumor with a sickle shape, which is unrealistic for a liver tumor. \textbf{(c)} A synthetic kidney tumor in the corticomedullary junction zone exhibiting a round nodular shape. However, this lesion fails to display a mass effect, as the healthy surrounding renal structure shows no deformation due to the tumor's inherent volume. Additionally, the texture and noise inside the tumor do not match the CT background. \textbf{(d)} A synthetic kidney tumor has a shape that matches the kidney "perfectly." However, solid tumors tend to grow expansively into a round nodular shape rather than precisely overlapping with the organ. \textbf{(e)} A synthetic tumor with a sickle shape, which is unrealistic for a pancreatic tumor. Additionally, the attenuation of this lesion is too low for a solid pancreatic tumor. \textbf{(f)} A synthetic pancreatic tumor adjacent to extra-pancreatic vessels. This tumor shows no mass effect, leaving the vessels without displacement or infiltration.
    } 
	\label{fig:supp_unrealistic_generation}
\end{figure}

Although \ourmodel\ is capable of generating highly realistic tumors, it also occasionally produces some that are less convincing. Consequently, about 50\% of the tumors are identified as inauthentic by the more experienced radiologist in the Visual Turing test. The tumors deemed inauthentic fall short in several aspects, such as shape, attenuation and noise distribution. Some synthetic tumors have inaccurate shapes, resembling flat, strip-like, or sickle-shaped lesions. In contrast, early-stage tumors originating from parenchymal organs typically exhibit a round or oval shape. Larger tumors fail to display a mass effect, characterized by the displacement of normal structures due to the tumor's inherent volume. Furthermore, some synthetic tumors inaccurately display a lower density, which is similar to that of fat or fluid. Finally, the noise distribution in some synthetic tumors does not match that in the CT background. We show several unrealistic generation cases in \figureautorefname~\ref{fig:supp_unrealistic_generation}.

\subsection{Challenging Cases Analysis}
\begin{figure}[h]
	\centering
	\includegraphics[width=\columnwidth]{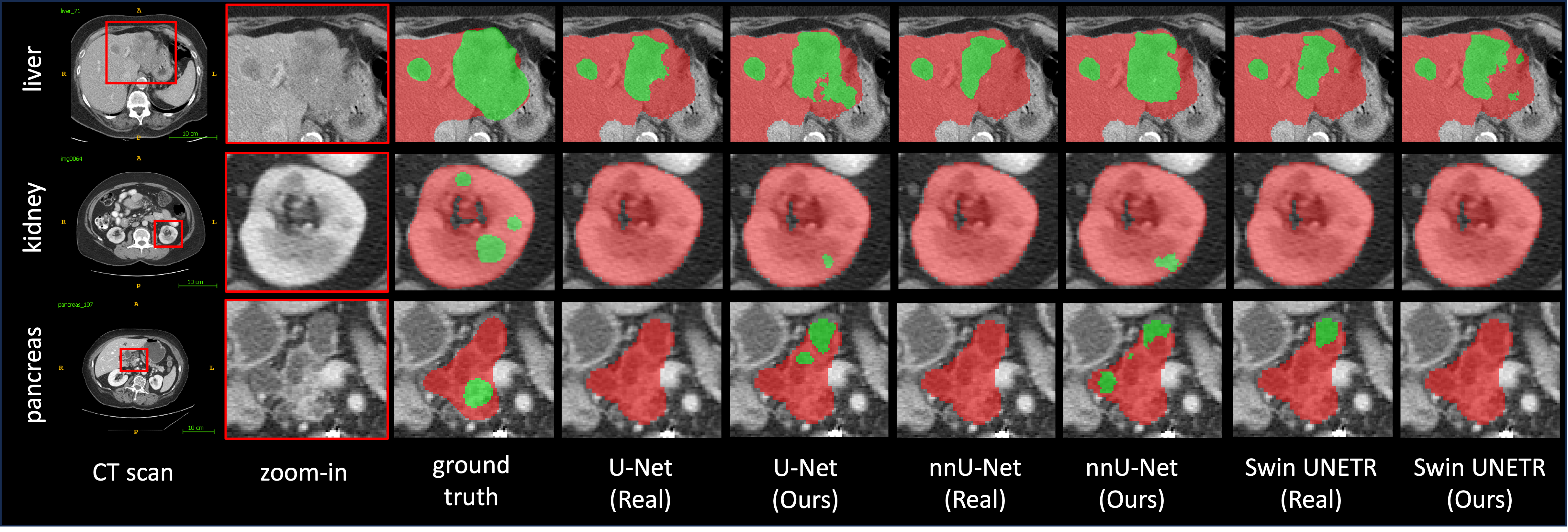}
    \caption{
    \textbf{Challenging cases for tumor segmentation.} \textbf{Liver tumor:} An irregular heterogeneous hypoattenuating mass is observed in the left lateral lobe of the liver, showing extracapsular penetration and infiltration into the stomach. The area with lower hypoattenuation inside the tumor indicates necrosis.
    \textbf{Kidney tumor:} This is a case with multiple renal tumors in the cortex and corticomedullary junction zone. Some of the tumors exhibit isoattenuation with the renal medulla, posing difficulties in detecting the lesions.
    \textbf{Pancreatic tumor:} In the illustrated case within this figure, the segmentation model erroneously classifies dilated pancreatic ducts as a tumor, resulting in inaccurate segmentation. Conversely, the genuine tumor, characterized by its indistinct boundaries, is not successfully segmented. 
    } 
	\label{fig:supp_challenging_cases}
\end{figure}

Instances of low performance are observed with all Segmentation Models (U-Net, nnU-Net, Swin UNETR) trained on both synthetic and real tumors. These instances often involve tumors with uncommon imaging features, as identified by experienced radiologists. Examples of these tumors from the liver, kidney, and pancreas are provided in \figureautorefname~\ref{fig:supp_challenging_cases}.

\end{document}